\documentclass[manuscript]{aastex}
\shorttitle{The X-ray and Optical Jets of Quasars 0106+013 \& 3C\,345}
\shortauthors{Kharb et al.}
\begin{document}
\title{Chandra \& HST Imaging of the Quasars PKS B0106+013 \& 3C\,345: Inverse Compton X-rays and Magnetized Jets} 
\author{P. Kharb}
\affil{Department of Physics, Rochester Institute of Technology, Rochester, NY 14623}
\email{kharb@cis.rit.edu}
\author{M. L. Lister}
\affil{Department of physics, Purdue University, West Lafayette, IN 47906}
\author{H. L. Marshall}
\affil{Center for Space Research, Massachusetts Institute of Technology, Cambridge, MA 02139}
\and
\author{B. S. Hogan}
\affil{Department of physics, Purdue University, West Lafayette, IN 47906}
\affil{El Camino College, 16007 Crenshaw Blvd. Torrance, CA 90506}
\affil{Marymount College, 30800 Palos Verdes Drive East, Rancho Palos Verdes, CA 90275}

\begin{abstract}
We present results from deep ($\sim$70 ks) {\it Chandra} ACIS observations and {\it Hubble Space Telescope} (HST) ACS F475W observations of two highly optically polarized quasars belonging to the MOJAVE (Monitoring Of Jets in Active galactic nuclei with VLBA Experiments) blazar sample, {\it viz.,} PKS B0106+013 and 1641+399 (3C\,345). These observations reveal X-ray and optical emission from the jets in both sources. X-ray emission is detected from the entire length of the 0106+013 radio jet, which shows clear bends or wiggles - the X-ray emission is brightest at the first prominent kpc jet bend. A picture of a helical kpc jet with the first kpc-scale bend representing a jet segment moving close(r) to our line of sight, and getting Doppler boosted at both radio and X-ray frequencies, is consistent with these observations. The X-ray emission from the jet end however peaks at about 0$\farcs$4 ($\sim$3.4 kpc) upstream of the radio hot spot. Optical emission is detected both at the X-ray jet termination peak and at the radio hot spot. The X-ray jet termination peak is found upstream of the radio hot spot by around 0$\farcs$2 ($\sim$1.3 kpc) in the short projected jet of 3C\,345. HST optical emission is seen in an arc-like structure coincident with the bright radio hot spot, which we propose is a sharp (apparent) jet bend instead of a terminal point, that crosses our line of sight and consequently has a higher Doppler beaming factor. A weak radio hot spot is indeed observed less than 1$\arcsec$ downstream of the bright radio hot spot, but has no optical or X-ray counterpart. {By making use of the pc-scale radio and the kpc-scale radio/X-ray data, we derive constraints on the jet Lorentz factors ($\Gamma_{jet}$) and inclination angles ($\theta$): for a constant jet speed from pc- to kpc-scales, we obtain a $\Gamma_{jet}$ of $\sim$70 for 0106+013, and $\sim$40 for 3C\,345. On relaxing this assumption, we derive a $\Gamma_{jet}$ of $\sim$2.5 for both the sources. Upper limits on $\theta$ of $\sim13\degr$ are obtained for the two quasars.} 
Broad-band (radio-optical-X-ray) spectral energy distribution modeling of individual jet components in both quasars suggests that the optical emission is from the synchrotron mechanism, while the X-rays are produced via the inverse Compton mechanism from relativistically boosted cosmic microwave background seed photons. The locations of the upstream X-ray termination peaks strongly suggest that the sites of bulk jet deceleration lie upstream (by a few kpc) of the radio hot spots in these quasars. These regions are also the sites of shocks {or magnetic field dissipation}, which reaccelerate charged particles and produce high energy optical and X-ray photons. This is consistent with the best fit SED modeling parameters of magnetic field strength and electron powerlaw indices being higher in the jet termination regions compared to the cores. The shocked jet regions upstream of the radio hot spots, the kpc-scale jet wiggles {and a ``nose cone'' like jet structure} in 0106+013, and the V-shaped radio structure in 3C\,345, are all broadly consistent with instabilities associated with Poynting flux dominated jets. A greater theoretical understanding and more sensitive numerical simulations of jets spanning parsec- to kpc-scales are needed, however, to make direct quantitative comparsions. 
\end{abstract}

\section{Introduction}
Blazars are radio-loud active galactic nuclei (AGNs) characterized by the presence of bright, compact radio cores that are highly polarized and variable in total and polarized intensity. They also typically exhibit superluminal motion in their parsec-scale jets. This extreme behavior is commonly understood to be a consequence of relativistic Doppler boosting effects in jets that are aligned at small angles to our line of sight \citep[e.g.,][]{Angel80}. In terms of emission-line properties, the blazar class comprises of optically violently variable (OVV) quasars and BL~Lac objects, with their rest-frame emission-line equivalent widths greater than or less than 5 \AA, respectively \citep{Stocke91,Stickel91}. In terms of radio jet properties, apart from total radio power, quasars and BL~Lacs are distinguished by the presence or absence of narrow collimated jets with terminal hot spots, respectively \citep[although see][]{Rector01,Kharb10}. 

In the last decade, the {\it Chandra} X-ray Observatory has added a new dimension to the field of jet astrophysics by detecting large kpc-scale X-ray jets in the majority of blazars studied \citep[e.g.,][]{Sambruna04,Marshall05,Marshall11,Hogan11}. This is intriguing because X-ray synchrotron emitting electrons are expected to have short radiative lifetimes, typically of the order of 10 years, in an ``equipartition'' magnetic field ($B_{eq}$, typically $\le$0.1 mG). While the X-ray emission in some low radio power jets can indeed be best described as synchrotron radiation \citep[e.g.,][]{Harris98,Sambruna07}, X-ray jets in powerful radio sources seem to be consistent with inverse-Compton (IC) emission associated with boosted cosmic microwave background (CMB) photons \citep[e.g.,][]{Tavecchio00,Celotti01}, or IC emission associated with the synchrotron photons themselves, {\it i.e.,} the synchrotron self-Compton (SSC) mechanism \citep[e.g.,][]{Hardcastle01}. A two-zone model has been invoked for the jet in the quasar 3C\,273, where a fast-moving ``spine'' produces most of the X-ray emission through the IC/CMB mechanism, and a slower-moving ``sheath'' produces X-rays through the synchrotron process \citep{Jester06}.
The IC/CMB model is attractive as it can explain the emission in a majority of X-ray jets, without straying far from the minimum energy condition or the ``equipartition'' magnetic field \citep{HarrisKrawczynski02}. However, the model requires that the jet remain highly relativistic on kiloparsec-scales (bulk Lorentz factors, $\Gamma_{jet} \sim 5-10$), so that in the jet frame, the CMB photon energy density is boosted by $\Gamma_{jet}^2$. The difficulty arises because a $\Gamma_{jet} \sim2$ is generally deduced for kpc-scale jets, through jet to counterjet brightness ratio arguments \citep{Bridle94,Wardle97}. However, it is fair to say that jet speeds in kpc jets are still largely unconstrained due to the lack of direct observations. The SSC model, on the other hand, usually requires the jet magnetic field to be significantly lower than $B_{eq}$. The synchrotron emission model is also inconsistent with the shape of the broad band spectral energy distributions (SEDs) in many quasar jet knots \citep{Harris06}. 
Therefore, the exact X-ray emission mechanism(s) in the kpc-scale jets or jet knots of different AGN subclasses, is still an unresolved issue. 

The uncertainties associated with the X-ray jet emission mechanisms are compounded by the fact that observations of large well-defined samples of distinct AGN subclasses have not yet been carried out. Only pilot observations of sizable samples ($>10$ sources) have been presented so far by \citet{Sambruna04,Hardcastle04,Marshall05,Massaro10,Marshall11}. We recently concluded a pilot ($\sim$10 ksec) X-ray study of 27 superluminal quasars belonging to the MOJAVE-{\it Chandra} sample \citep[MCS,][]{Hogan11}. The MOJAVE (Monitoring Of Jets in Active galactic nuclei with VLBA Experiments) sample is a complete parsec-scale flux density-limited sample at 15\,GHz of 135 radio-loud AGNs, most of which are blazars \citep{ListerHoman05}. The sample members of the MCS have large 1.4 GHz Very Large Array (VLA) extended flux densities ($>100$ mJy), and long radio jets ($\ge3''$ in extent). The pilot study revealed X-ray jets in nearly 80\% of the sample \citep{Hogan11}. As the majority of the MCS quasars lie between redshifts 0.5 and 1 (three sources are at $z>$1), the high detection rate could be considered to be consistent with the IC/CMB model, as the CMB photon energy density increases with redshift as $(1+z)^{4}$. 

In order to expressly test this idea, by creating and modeling broad band SEDs of individual jet components, we recently acquired deep $\sim$70 ksec {\it Chandra} and single-orbit {\it Hubble Space Telescope} (HST) imaging data on two MCS quasars, $viz.,$ PKS B0106$+$013, and 1641$+$399 (3C\,345). The results from these observations are presented here.  The paper is structured as follows. Section \ref{secsources} describes the two targets, while Section \ref{secobsvn} describes the multi-waveband observations and data-analysis. Section \ref{secspectral} discusses the imaging, X-ray spectral analysis and broad-band SED modeling. {Section \ref{secK} describes an attempt to constrain the jet speed and viewing angle through pc- and kpc-scale data.} The discussion and summary follow in Section \ref{secdiscussion}. The spectral index $\alpha$ is defined such that flux density at frequency $\nu$ is, $S_{\nu}\propto\nu^{-\alpha}$ and the photon index $\Gamma=1+\alpha$. Throughout this paper, we adopt the cosmology in which $H_0$=71 km s$^{-1}$ Mpc$^{-1}$, $\Omega_m$=0.27 and $\Omega_{\Lambda}$=0.73. 

\section{Observed Sources} 
\label{secsources}
For our deep follow-up {\it Chandra}\,-\,HST study, we chose two MCS quasars which showed X-ray jets in the pilot observations, had large 1.4 GHz VLA extended flux densities ($>500$ mJy), and archival HST imaging data through at least one broad band filter.

\subsection{PKS B0106+013}
PKS B0106+013 is a highly optically polarized quasar (HPQ) at a redshift of 2.099 (1$\arcsec$ at the distance of the source corresponds to a spatial scale of $\sim$8.4 kpc). Its projected $\sim$38 kpc radio jet is clearly one-sided (Figure \ref{figxrayimage}), which could be suggestive of a highly relativistic jet on kpc-scales, oriented close to our line of sight. The kpc jet has a gently undulating S-shape around a position angle of 180$\degr$. Diffuse emission from a radio lobe is not detected in its 1.6\,GHz or 4.9\,GHz images, either in the jet or the counterjet direction \citep{Cooper07}. This may be due to the fact that only the VLA A-array configuration was used at both the radio frequencies, and could therefore be missing a significant fraction ($\sim$15\%$-$20\%) of the diffuse lobe emission as opposed to combined array observations \citep[see][]{Kharb08,Kharb10}. Indeed, the averaged total flux density at 4.8 GHz from the single-dish University of Michigan Radio Astronomy Database (UMRAO) is $\sim$3.0 Jy for the year 1988, while it is $\sim$2.5 Jy from our January 1989 VLA data, indicating a $\sim17\%$ loss of (possibly extended) emission. {However, there is an alternate explanation which is more in line with our conclusions below that this jet is highly magnetized. This quasar may represent a rare class of powerful radio sources that do not possess radio lobes, but rather have ``nose-cones''. Magnetized plasma then moves forward through the terminal shocks in the jets, rather than backward like in radio lobes \citep{Komissarov99,Clarke86}.} 

The parsec-scale 15\,GHz radio emission of 0106+013 has been monitored by MOJAVE \citep{Lister09} and the 2 cm Survey \citep{Kellermann04} since 1995. The jet structure is dominated by a bright, optically thick core, and at least four individual jet features have been seen to move outward from the core region \citep{Lister09b}. These have apparent speeds ranging from  21$c$ to 26$c$, which are within the upper range seen in the radio-selected MOJAVE jet sample, and imply a bulk Lorentz factor of at least $\sim21$. The features are moving outward at different individual position angles with respect to the core (between 229$\degr$ and 260$\degr$), and are following shallow curved paths. The jet rapidly drops below the sensitivity level of the MOJAVE 2 cm images at $\sim$3 milliarcsec (mas) from the core. VLBA images at longer wavelengths \citep{Fomalont00} indicate a structure that curves southward  out to $\sim30$ mas at a position angle of 210$\degr$. 

\subsection{3C\,345}
\begin{figure}
\centering{
\includegraphics[width=8.6cm]{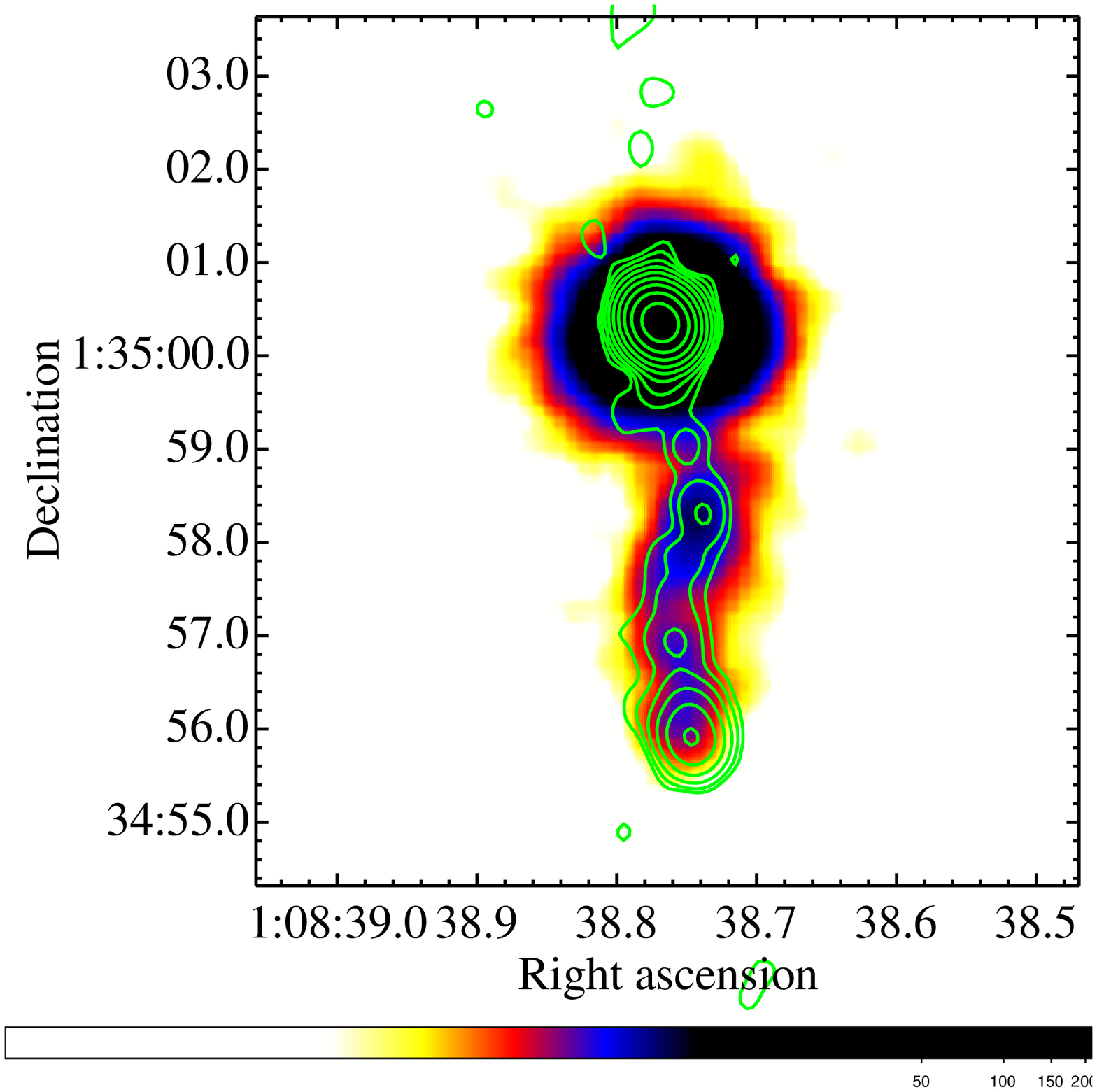}
\includegraphics[width=8.6cm]{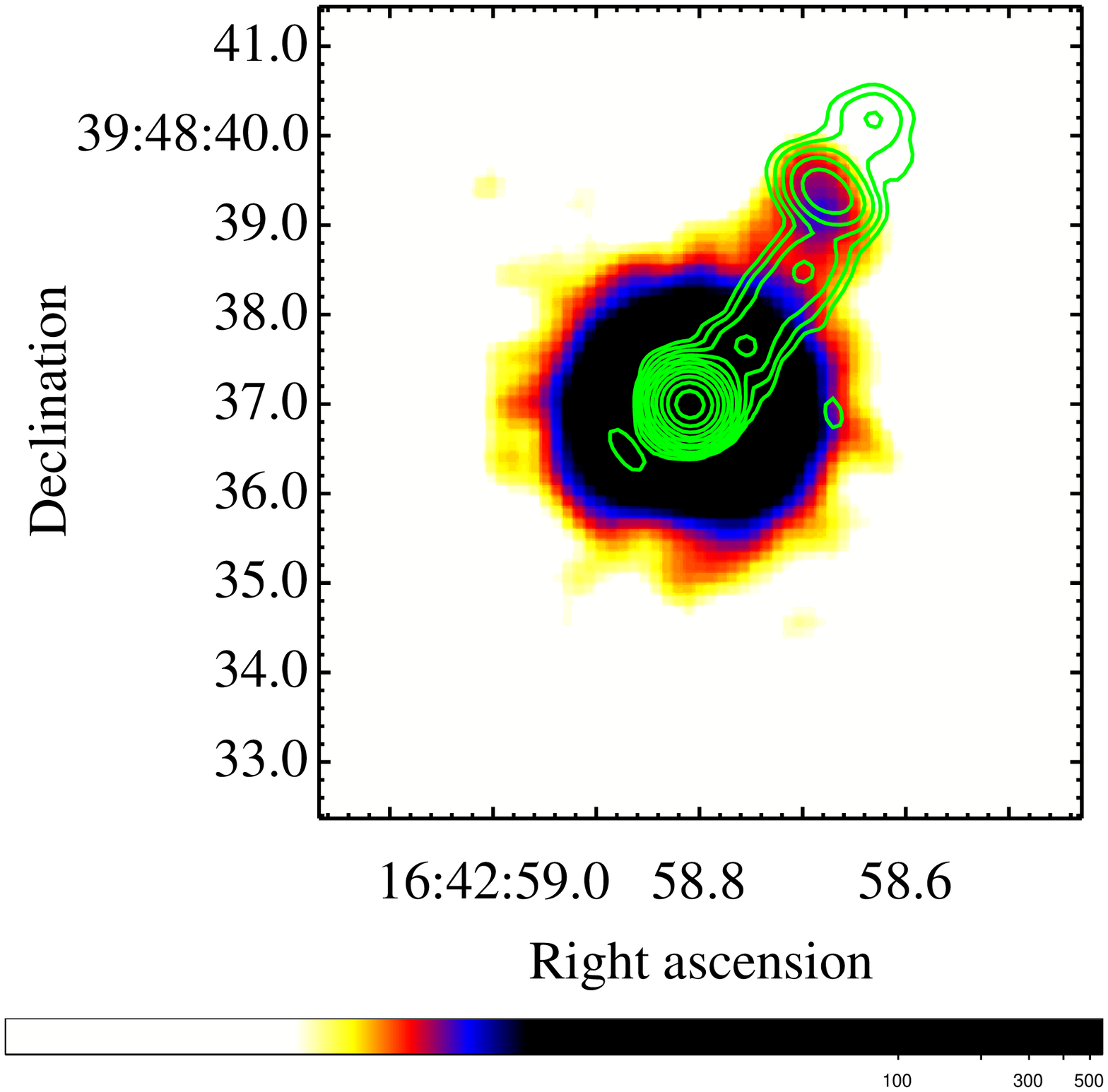}}
\caption{\small Energy filtered (0.5$-$7.0 keV) {\it Chandra} X-ray image of 0106+013 (top) and 3C\,345 (bottom) in color, with the 4.9 GHz VLA radio contours superimposed. Images were binned to 0.25 of the native pixel size, and smoothed with a Gaussian of kernel radius = 3. The radio contours are (top) 0.001, 0.002, 0.006, 0.01, 0.03, 0.08, 0.2, 0.5, 1.3, 3.2, and (bottom) 0.002, 0.005, 0.01, 0.02, 0.05, 0.11, 0.24, 0.52, 1.14, 2.5, 5.5 Jy~beam$^{-1}$, respectively. The color scale goes from white to black for the X-ray intensity range of 0.5 to 70 counts~arcsec$^{-2}$.}
\label{figxrayimage}
\end{figure}

3C\,345 or B3 1641+399 is a powerful HPQ located at redshift $z = 0.593$ (1$\arcsec$ at the distance of the source corresponds to $\sim$6.6 kpc). The primary radio structure consists of a short jet at position angle $327\degr$ that contains a bright hotspot, 2$\farcs$7 away from the unresolved core (Figure \ref{figxrayimage}). The radio jet extends to a projected distance of $\sim$25 kpc. The 1.5\,GHz image shows an asymmetric halo centered on the core that has a radius of approximately 10$\arcsec$. Although there is no clear indication of a radio counterjet, there is a suggestion of a weak hot spot along the counterjet direction so that the hotspot-core-counter-hotspot forms a V-shape \citep[see][]{Murphy93}. This is likely to be the result of a somewhat misaligned jet and counterjet that is oriented at a very small angle to line of sight. We note that the averaged total flux density at 4.8 GHz from December, 1988 to February, 1989, is $\sim$8.6 Jy from the single-dish UMRAO database, while it is $\sim$8.8 Jy from our January 1989 VLA data. This indicates that the VLA 4.8 GHz A-array observations do not miss much/any diffuse extended emission in this source. Core variability and/or amplitude and phase calibration errors could be contributing factors in the flux differences. It is interesting to note that V-shaped sources and S-shaped jets (as in 0106+013) are easier to explain in Poynting flux or magnetically-dominated jets, as they are easier to deflect in the presence of clumpy media \citep{Benford78,Nakamura07}. We return to this suggestion in Section \ref{secdiscussion}.

Being one of the most compact and powerful radio sources in the northern sky, the parsec-scale radio jet of 3C\,345 has been studied for several decades with Very Long Baseline Interferometry. The jet starts out at a position angle of $\sim270\degr$, but then at 4 mas from the core it abruptly curves northward and then propagates at a position angle of $\sim 315\degr$. A long term campaign to study the jet kinematics \citep{Schinzel10} has shown that this bend has slowly migrated outward over time, and they have found numerous superluminal features with complex motions on scales of 0.3 to 15 mas. There appears to be no common velocity for features within 0.7 mas of the core, but at farther distances numerous features have been seen to move at speeds between 10$c$ and 13$c$. The MOJAVE program has reported a maximum speed of 19.7$c$ in the jet \citep{Lister09b}. Many of the features have helical or strongly curved trajectories on the sky, leading \citet{Lobanov05} to postulate a precessing jet model driven by a supermassive binary black hole system in 3C\,345.

\begin{figure}
\centering{
\includegraphics[width=8.8cm]{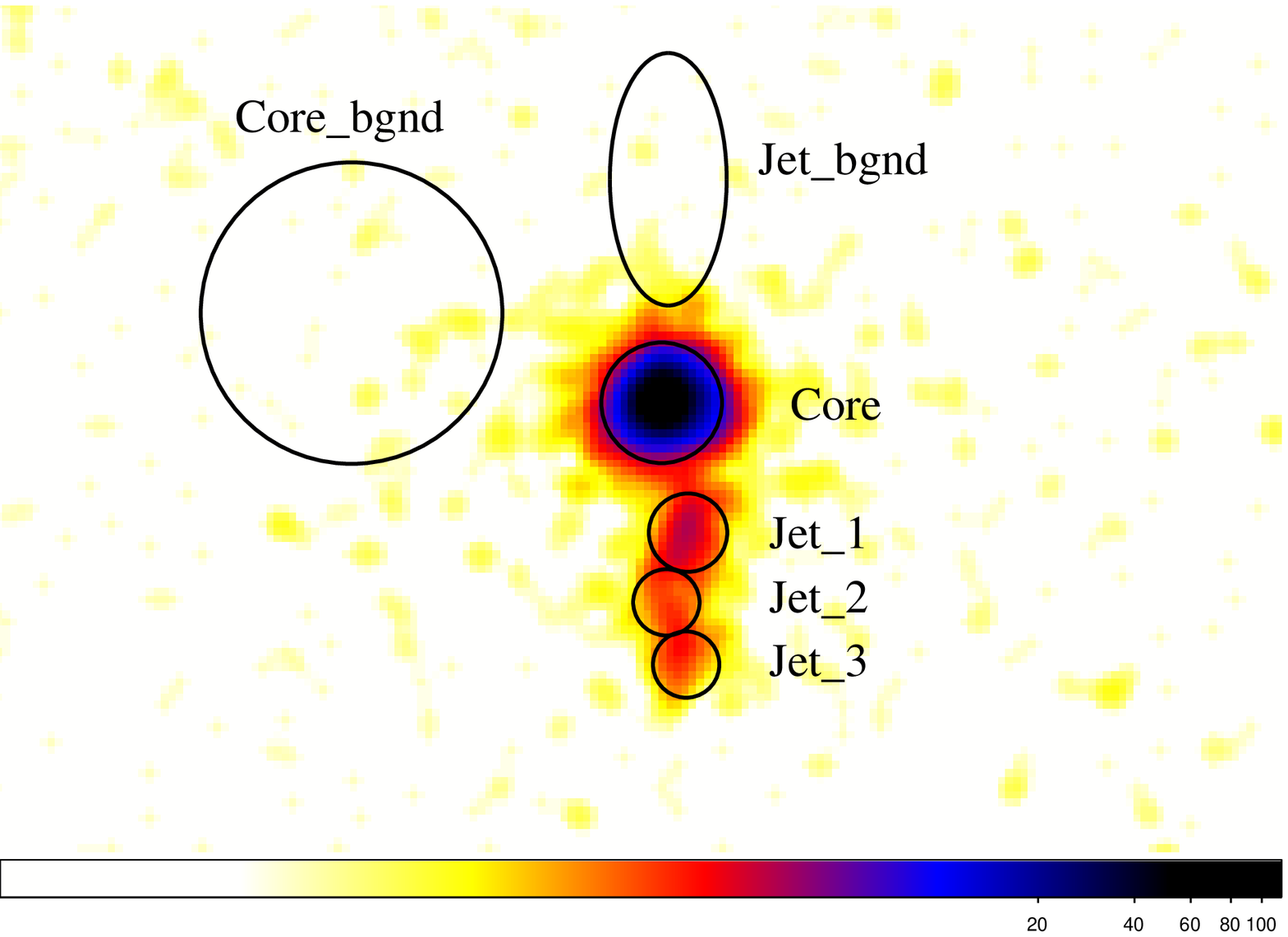}
\includegraphics[width=8.8cm]{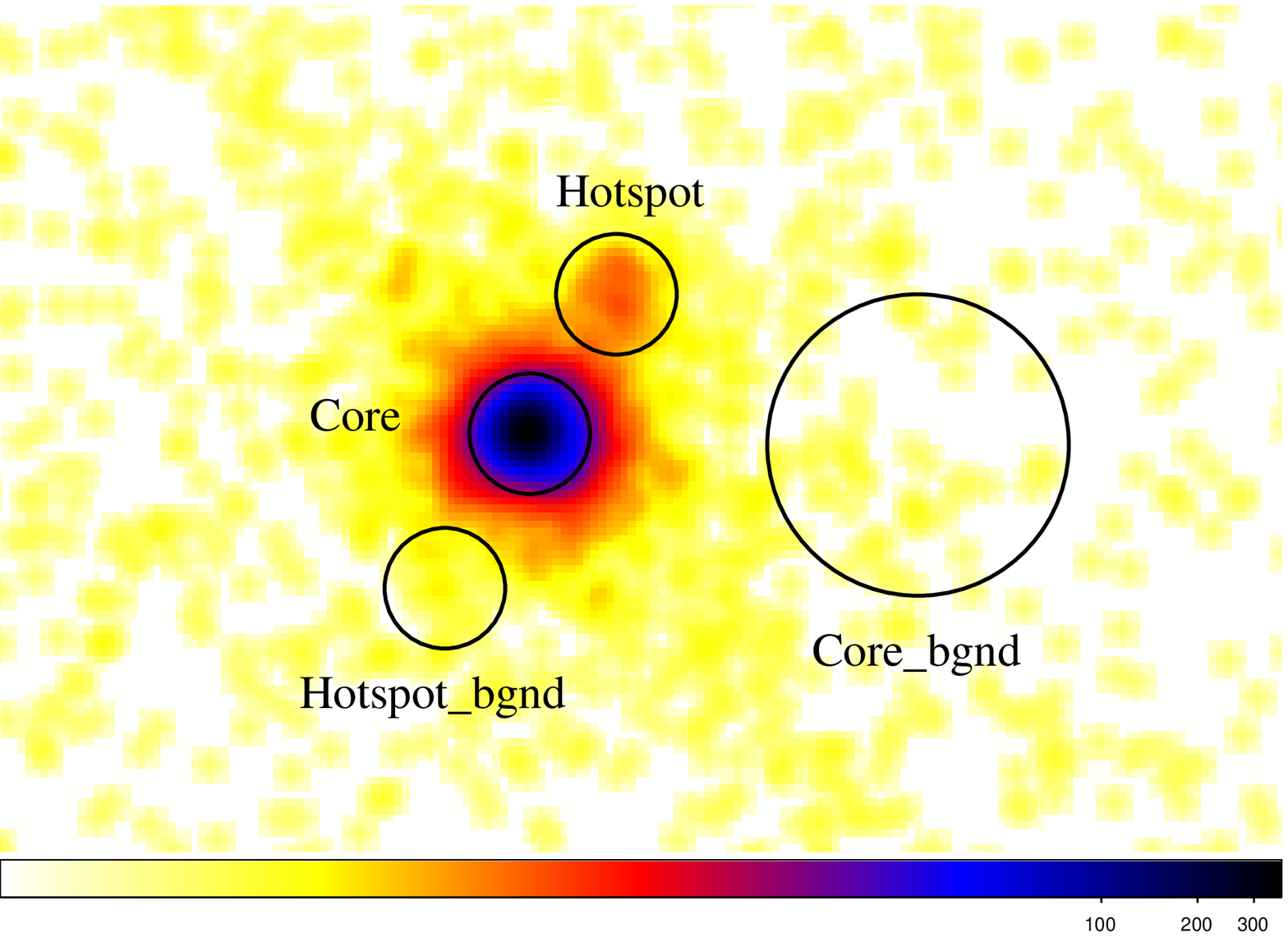}}
\caption{\small Regions used to extract the X-ray spectra for the cores, jet components, and background for 0106+013 (top panel) and 3C\,345 (bottom panel).}
\label{figregions}
\end{figure}

\section{Multiwavelength Observations and Data Analysis}
\label{secobsvn}

\subsection{Chandra Data}
The observations for 0106+013 took place on 2008 November 7 and 9, and for 3C\,345 on 2009 November 1 and 3 (Table \ref{tabderived}). The observations were obtained using the AXAF CCD Imaging Spectrometer (ACIS) S3 chip (which is back illuminated for low-energy response) in the very faint (VFAINT) timed mode. In order to reduce the effect of pileup of the bright quasar cores, the observations were carried out using the 1/8 subarray mode (frame time = 0.441 sec). For both the sources a range of roll angles were specified so that the charge transfer trail of the nucleus did not contaminate the jet emission.

The {\it Chandra} data were reduced and analysed using the CIAO software version 4.2 and calibration database (CALDB) version 4.2.1\footnote{http://cxc.harvard.edu/ciao/}. Using the CIAO tool ``$acis\_process\_events$'' we reprocessed the level 1 data to remove the 0.5 pixel randomization. We did not attempt the VFAINT mode background cleaning since both the targets were affected by pileup in the nuclear regions \citep[e.g.,][]{Davis01}. The observations reveal that the core of 0106+013 has 0.06~counts~frame$^{-1}$, corresponding to a small pileup fraction (2\%), while the core of 3C\,345 has 0.18~counts~frame$^{-1}$, corresponding to a pileup fraction of 8\%. The data were filtered on the Advanced Satellite for Cosmology and Astrophysics (ASCA) grades 0, 2, 3, 4, 6 and on status. Finally, the Good Time Intervals (GTI) supplied with the pipeline were applied to create the new level 2 events file. The background light curves were examined for flares, using the relevant thread in CIAO\footnote{http://cxc.harvard.edu/ciao/threads/filter/index.sl.html\#filterback}. A binning time of 100$\times$frame-time was used for creating the light curves. No significant background flares were detected in the data of either source. 

The X-ray images with the 4.9 GHz radio contours superimposed are displayed in Figure \ref{figxrayimage}. For the purpose of creating these images, we merged the two observations of each source with ``$dmmerge$'' in CIAO, after reprojecting the two new event 2 files around the tangent point of one of them with CIAO tool ``$reproject$\_$events$''. The X-ray and radio cores were matched using the FITS file editing software, Fv\footnote{http://heasarc.nasa.gov/ftools/fv/}, after first ``locking'' the crosshair on the radio core World Coordinate System (WCS) position in the astronomical imaging and data visualization application, DS9, and noting the position difference in pixels with respect to the X-ray core. The X-ray image was smoothed with a Gaussian of kernel radius 3 prior to the estimation of the pixel shift. The final shifts in the x, y direction were 0.24, 0.25 pixels for 0106+013, and 0.60, 0.16 pixels for 3C\,345. These pixel shifts are consistent with the expected accuracy of the {\it Chandra} astrometry. Table \ref{tabderived} lists some observed and derived source parameters. The magnetic field strength, B$_{eq}$, was derived assuming an ``equipartition'' of relativistic particle energy density and magnetic field energy density \citep{Burbidge59}, for a cylindrical jet geometry. The radio spectrum was assumed to extend from 10$^{7}$ to 10$^{15}$ Hz, with a spectral index of 0.8. Furthermore, a volume filling factor of unity, and zero contribution to the particle energy density from protons was assumed.

The CIAO tool ``{\it specextract}'' was used to extract the spectrum from the core and jet regions. Figure \ref{figregions} displays the regions used in the spectral extraction. Subsequently, the spectral fitting and analysis was done using the XSPEC software ({\it HEASOFT version 6.6.3}), using events with energy $>$0.5 keV and $<$10 keV, where the calibration is well known. We attempted the spectral fitting for the core, using a circular region with diameter of 2$\arcsec$, centered on the pixel with the highest number of counts. The background was chosen as a source-free circular region with a diameter of 5$\arcsec$.  A binning of 50 counts\,bin$^{-1}$ was used for the core. 

We extracted jet spectra in 0106+013 from three circular regions of diameter 1$\farcs$3 (jet$\_$1), 1$\farcs$1 (jet$\_$2), and 1$\farcs$1 (jet$\_$3), respectively (Figure \ref{figregions}, top panel). The background region for the jet spectra was chosen as an elliptical region of size 2$\arcsec \times 4\arcsec$. A binning of 20 counts\,bin$^{-1}$ was used for the jet region ``jet$\_$1'', and 15 counts\,bin$^{-1}$ for regions ``jet$\_$2'' and ``jet$\_$3''. The jet knot/hot spot spectra for 3C\,345 were extracted using a circular region of diameter 2$\arcsec$, with a similar sized region used for the jet knot/hot spot background spectra (Figure \ref{figregions}, bottom panel). A binning of 20 counts\,bin$^{-1}$ was used for this region. Spectra from the two observations of each source (shown as red and black data in Figures \ref{figspectral0106} to \ref{figspectral1641}), were fit simultaneously in XSPEC. The results of the spectral analysis are presented in Table \ref{tabxspec} and discussed below in Section \ref{secxspec}.

\begin{deluxetable}{lll}
\tablecaption{Observed and Derived Source Parameters}
\tabletypesize{}
\tablewidth{0pt}
\tablehead{\colhead{Parameter}&\colhead{0106+013}&\colhead{1641+399}}
\startdata
OBSID & 10380, 10799 & 10379, 12011 \\
Exposure (sec) &33727, 31907 & 32770, 24865\\
Date & 2008-11-07 & 2009-10-27 \\
Jet P.A. (deg) &  185 & $-$30 \\
r$_{in}$, r$_{out}$ (arcsec) & 1.0, 5.0 & 1.0, 5.0 \\
S$_{4.9}$ (mJy) & 150.5 $\pm$ 0.8 & 309.8 $\pm$ 2.1\\ 
Count rate (ksec$^{-1}$) & 20.81 $\pm$  0.98  & 7.45 $\pm$  1.04 \\
S$_{1 keV}$ (nJy) & 20.8 & 7.4 \\
$\alpha_{rx}$ & 0.89 $\pm$ 0.00 & 0.99 $\pm$ 0.01 \\
R (IC/syn) &  197.4 &  34.3 \\
Volume ($pc^{-3}$)& 1.0E+12 & 5.2E+11\\
B$_{eq}$ (mG)& 157 & 96 \\
K & 24$\pm$5 & 23$\pm$4
\enddata
\vspace{-0.6cm}
\tablecomments{r$_{in}$, r$_{out}$ are the inner and outer extents of the jet region used for the derived parameters listed below. ``S$_{4.9}$'' and ``S$_{1 keV}$'' refer to the flux density at 4.9 GHz and 1 keV, respectively. ``R'' is the ratio of the observed X-ray (assumed to be inverse Compton) and radio (synchrotron) emission in the jets. ``R'' along with the equipartition magnetic field, B$_{eq}$, is used for the estimation of the dimensionless ``K'' parameter ({cf.} equation 2). The K parameter can be used to provide constraints on the jet viewing angle and Lorentz factors, as described in Section \ref{secK}.}
\label{tabderived}
\end{deluxetable}

\begin{figure}
\centering{
\includegraphics[width=9.1cm]{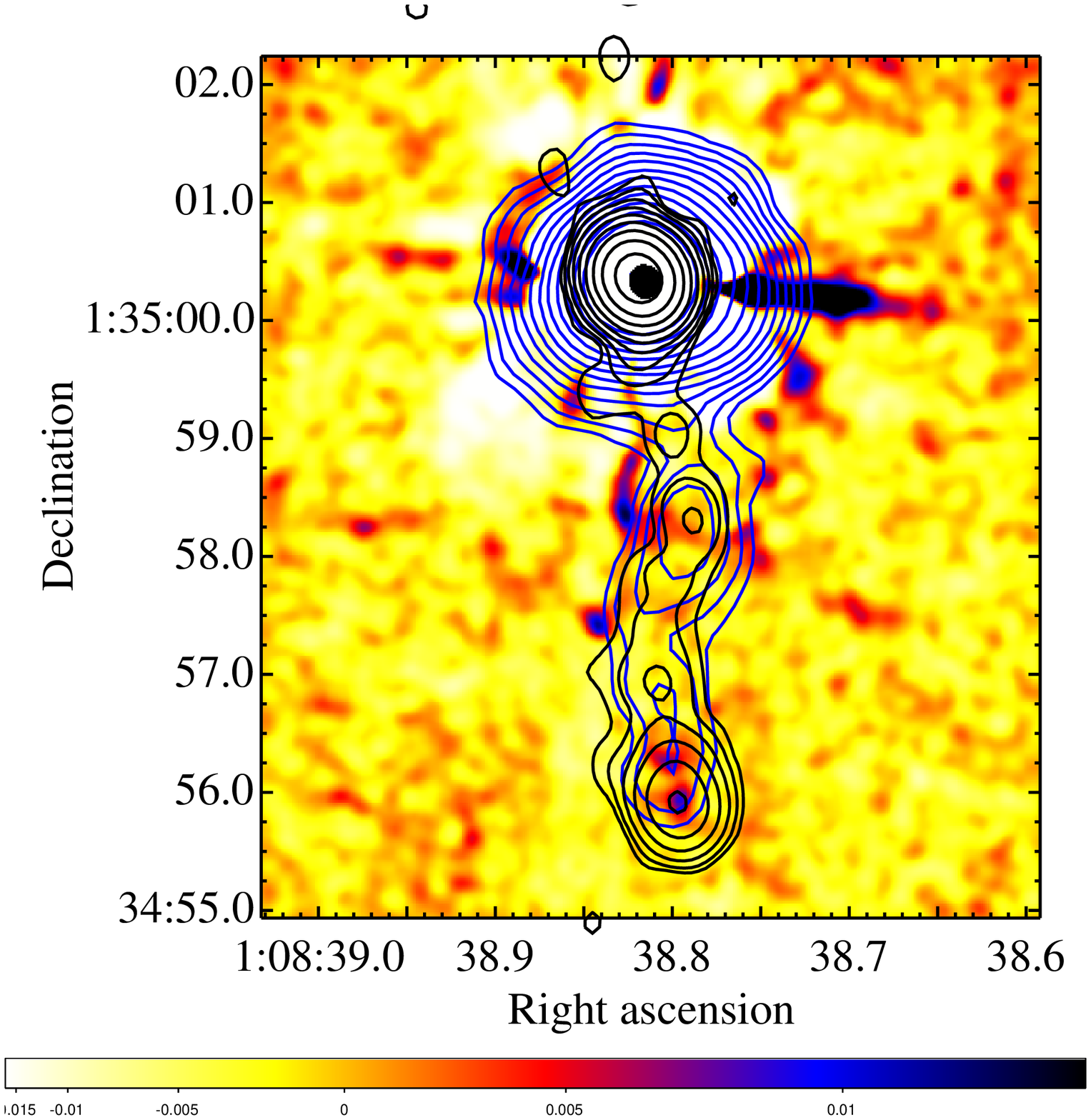}
\includegraphics[width=9cm]{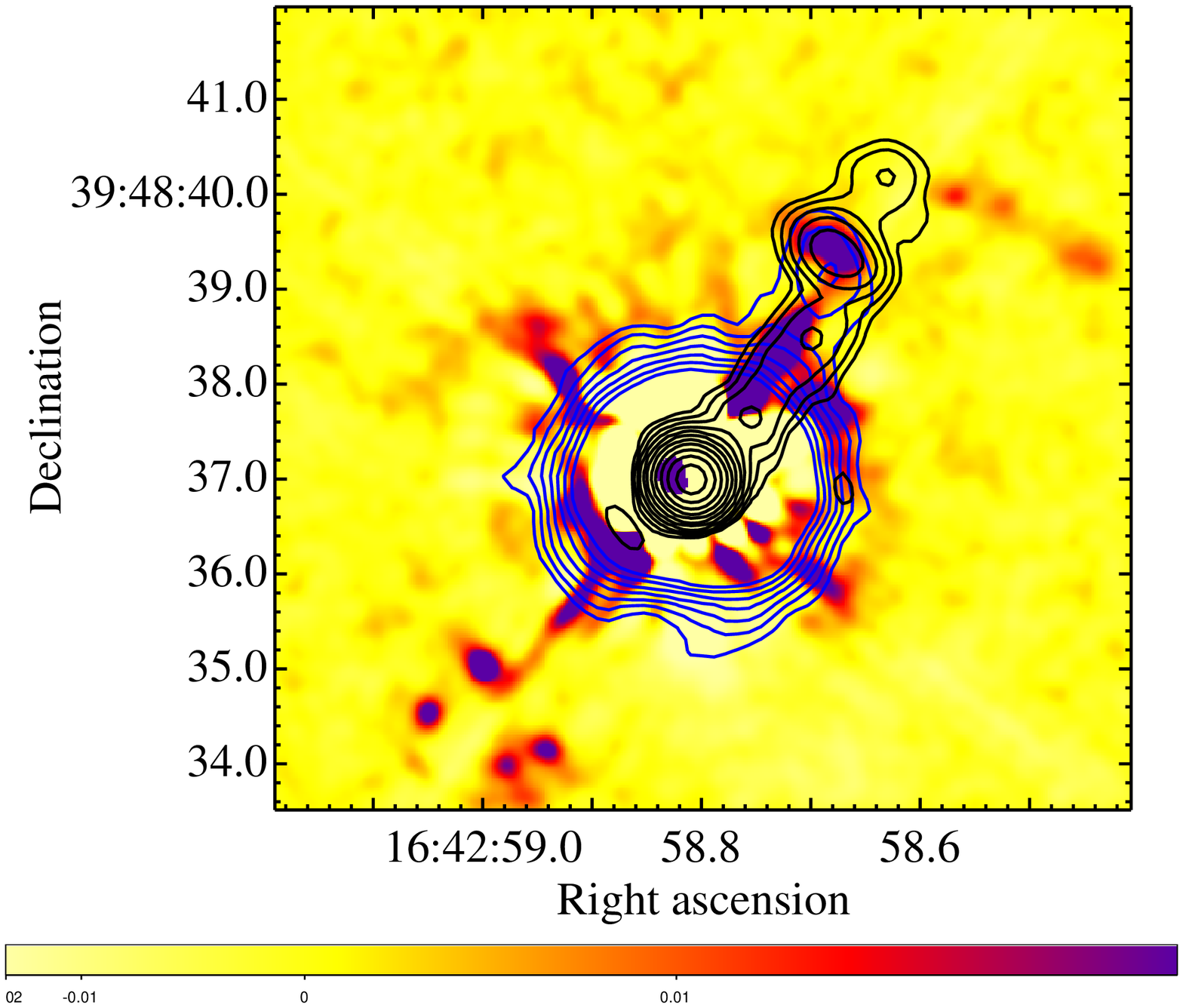}}
\caption{\small The PSF-subtracted HST/ACS F475W color image of 0106+013 (top) and 3C\,345 (bottom), with the 4.9 GHz radio (black) and X-ray (blue) contours superimposed. The HST image was smoothed with a Gaussian kernel radius of 9. The ``squared-zscale'' option was used for the color scale. The radio contour levels are the same as in Figure \ref{figxrayimage}. The X-ray contour levels are (top) 1.38, 1.98, 2.85, 4.10, 5.90, 8.48, 12.19, 17.54, 25.22, 36.27, 52.15, 75.00, and (bottom) 1.33, 1.77, 2.35, 3.12, 4.14, 5.5, 7.3 counts~arcsec$^{-2}$.}
\label{fighst}
\end{figure}
 
\subsection{HST Data} 
The {HST} observations for 3C\,345 took place on 2010 July 11, and for 0106+013 on 2010 November 2, with the Wide Field Channel \#1 of the Advanced Camera for Surveys (ACS/WFC1) through the F475W filter (pivot wavelength 4747 \AA). Each source was observed for a single {HST} orbit, which resulted in an exposure time of 2300 sec for 0106+013, and 2280 sec for 3C\,345. The data were obtained from the Space Telescope Science Institute (STScI) archive with the pipeline processing applied. The pipeline images had been fully calibrated with the MultiDrizzle program \citep{Koekemoer02}.
 
Due to an error in the {\it alternate} roll angle range specification for 3C\,345, a point spread function (PSF) spike fell close to the jet position. This has probably led to some contamination of the flux of the single bright jet knot/hot spot present in this source. As this source had archival data from the Space Telescope Imaging Spectrograph (STIS) CCD in the CLEAR (unfiltered) imaging mode \citep[see][]{Sambruna04}, we could obtain independent flux estimates for this jet feature. For this we combined the four sub-exposures in the STIS observations (at a pivot wavelength of 5850 \AA) using the NOAO Image Reduction Analysis Facility (IRAF) tool $crrej$, which also got rid of the cosmic rays. We find that using the largely un-contaminated STIS flux and the contaminated ACS flux, yields a relatively flat optical spectral index, $\sim0.06$ for the ``hot spot'' of 3C\,345. This suggests that the contamination from the spike may not be as significant as initially presumed. 

Although the HST archive lists wide-field planetary camera 2 (WFPC2) data through the F702W filter on 0106+013, the source is situated at the edge and just off the field of view. This makes the F475W image presented in this paper, the first {HST} image of 0106+013. 
Figure \ref{fighst} displays the PSF-subtracted {HST} images of 0106+013 and 3C\,345 with the 4.9~GHz radio and X-ray contours superimposed. The peak optical, X-ray and radio core positions were matched using the FITS file editing Fv software. We subtracted the scaled PSF from the image to reveal the jet features clearly. The PSFs were created using the TinyTim software\footnote{http://www.stecf.org/instruments/TinyTim/}. 
The scaling was done using the integrated flux of the quasar cores, rather than the peak flux. 
Figure \ref{fighst} highlights the result that in 0106+013, optical hot spots are spatially coincident with the radio and X-ray hot spots, while in 3C\,345, the optical hot spot is spatially coincident only with the radio hot spot.

The optical fluxes for the core and jet components of 0106+013 were estimated using circular regions of diameter $\sim0\farcs5$ in DS9 (Table \ref{tabmulti}). Source-free circular background regions of diameter 2$\arcsec$ were used for the jet components, while an annular region of diameter between 0$\farcs5-1\arcsec$ was used for the core. The core flux of 3C\,345 was estimated using similar source and background regions. The jet knot/hot spot optical flux for 3C\,345 was estimated using a circular region of diameter $\sim0\farcs66$ in DS9, while the background was estimated from an annular region between diameters $0\farcs66-1\farcs3$. The source counts in the respective regions were estimated using the Chandra Education Analysis Tool in DS9 through the Virtual Observatory option. The net background-subtracted counts were estimated using the relation, 
$Net~counts = Src~counts - (A_{Src}/A_{Bgnd}) \times Bgnd~counts$,
where ``A" is the area in pixels from which the counts were extracted, ``Src" and ``Bgnd'' refer to the source and background, respectively. Finally, the net counts were converted into fluxes using the inverse sensitivity keyword PHOTFLAM and the pivot wavelength keyword PHOTPLAM in the image headers. 
After experimenting with slightly different aperture sizes for the source and background regions, and noting the differences in the count estimates, we conclude that the errors in the optical flux estimations are typically $\sim5\%-10\%$ for the ACS measurement of 0106+013 and 3C\,345, and between $\sim5\%-20\%$ for the STIS measurement of 3C\,345. The symbol sizes in Figures \ref{figsed0106core} - \ref{figsed1641jet} can therefore be considered representative of the error bars in the optical flux estimates.

\begin{deluxetable}{ccccccccc}
\tablecaption{VLA A-Configuration Radio Observations}
\tablewidth{0pt}
\tablehead{
\colhead{Source}&\colhead{Obs}&\colhead{Freq}&\colhead{Project}&\colhead{TOS}&\colhead{Beam}&\colhead{rms}\\
\colhead{}&\colhead{Date}&\colhead{(GHz)}&\colhead{ID}&\colhead{(min)}&\colhead{($\arcsec\times\arcsec$)}&\colhead{(mJy~beam$^{-1}$)} }
\startdata
0106+013 &1989, Jan 07& 1.64 & AR197 & 115 &1.25$\times$1.14& 0.2 \\
                    &1989, Jan 07& 4.86 & AR197 & 108 &0.43$\times$0.38& 0.1 \\
1641+399 &1984, Dec 23& 1.55 & AB310 & 12   &1.36$\times$1.25& 0.8 \\
                    &1989, Jan 08& 4.86 & AR196 & 190 &0.36$\times$0.35& 0.25 
\enddata
\tablecomments{\small Cols.\,2, 3 \& 4: Observing date, frequency and project IDs. Col.\,5: Time on source. Col.\,6: Restoring beam-size. Col.\,7: Final {\it rms} noise in the radio images.}
\label{tabradio}
\end{deluxetable}

\subsection{VLA Data} 
The 1.6 and 4.9 GHz VLA A-array configuration data for 0106+013 and 3C\,345 were obtained from the NRAO data archive (Table \ref{tabradio}). Typically 10 min scans of the targets were interspersed with 3$-$4 min scans of the phase calibrators (2.5 min scans of target with 2 min scans of phase calibrator for the 1.6\,GHz observations of 3C\,345). The data reduction was carried out following standard calibration and reduction procedures in the Astronomical Image Processing System (AIPS). After the initial amplitude and phase calibration, AIPS tasks CALIB and IMAGR were used iteratively to self-calibrate \citep{Schwab80} and image the sources. The jet component flux densities were obtained in AIPS using verbs TVWIN+IMSTAT, while the core estimates were derived using the Gaussian-fitting task JMFIT. 
For the purpose of creating the broad-band SEDs, we created higher resolution 1.6 GHz radio images of beam-sizes $\sim0\farcs7$ (by fixing the BMAJ and BMIN parameters in IMAGR) to approximately match the resolution of the 4.9 GHz radio, the optical and X-ray images, before estimating the flux densities. The flux estimates are listed in Table \ref{tabmulti}.

\begin{figure}
\centering{
\includegraphics[width=16.5cm]{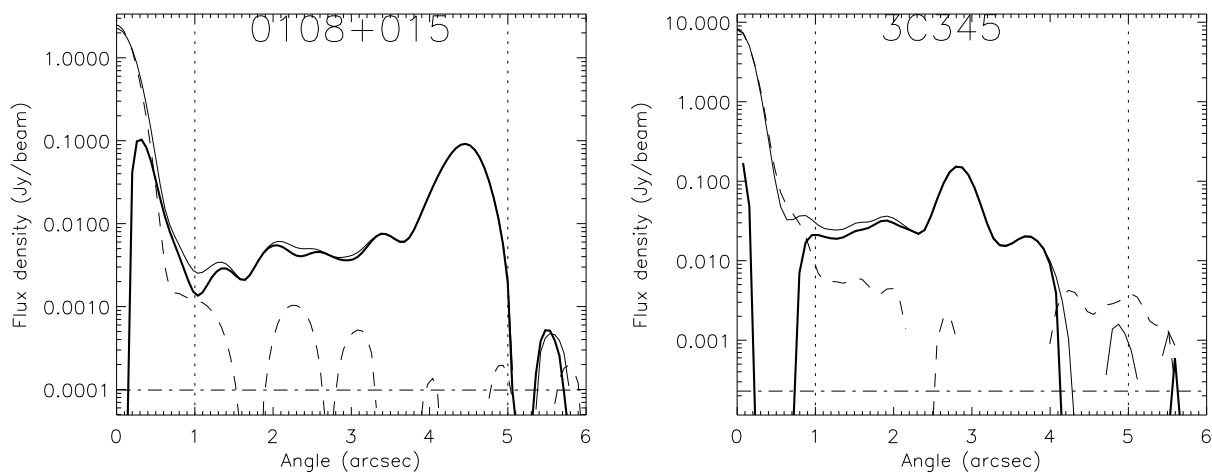}}
\vspace{-14cm}
\caption{\small Surface brightness profiles of the radio jet at 4.9\,GHz for 0106+013 (left panel) and 3C\,345 (right panel). The thin solid lines give the radio profiles along the position angle of the jets. The dashed lines indicate the radio profile at a position angle of 90$\degr$ counterclockwise from the jet to avoid any non-jet emission and counterjet emission. The solid, bold line indicates the difference between the two profiles so that core emission is removed and the effective flux can be measured. The horizontal dot-dashed lines are set to a value five times the average noise level and the vertical dashed lines show the inner and outer radius limits.}
\label{figradprofile}
\end{figure}
\begin{figure}
\centering{
\includegraphics[width=16.5cm]{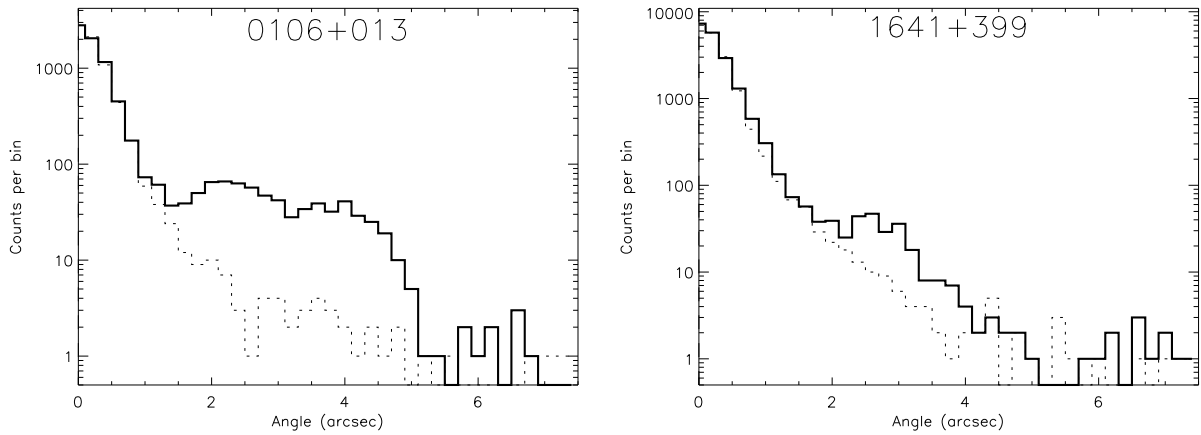}}
\vspace{-11.5cm}
\caption{\small X-ray jet profiles for 0106+013 (left panel) and 3C\,345 (right panel). These are represented as histograms of the counts in 0$\farcs$2 bins. The solid lines give the profile along the position angle of the jet, as defined by the radio images. The dashed lines show the profile along the counterjet direction, which is defined as 180$\degr$ opposite to the jet.}
\label{figxrayprofile}
\end{figure}
\begin{figure}
\centering{
\includegraphics[width=7.5cm]{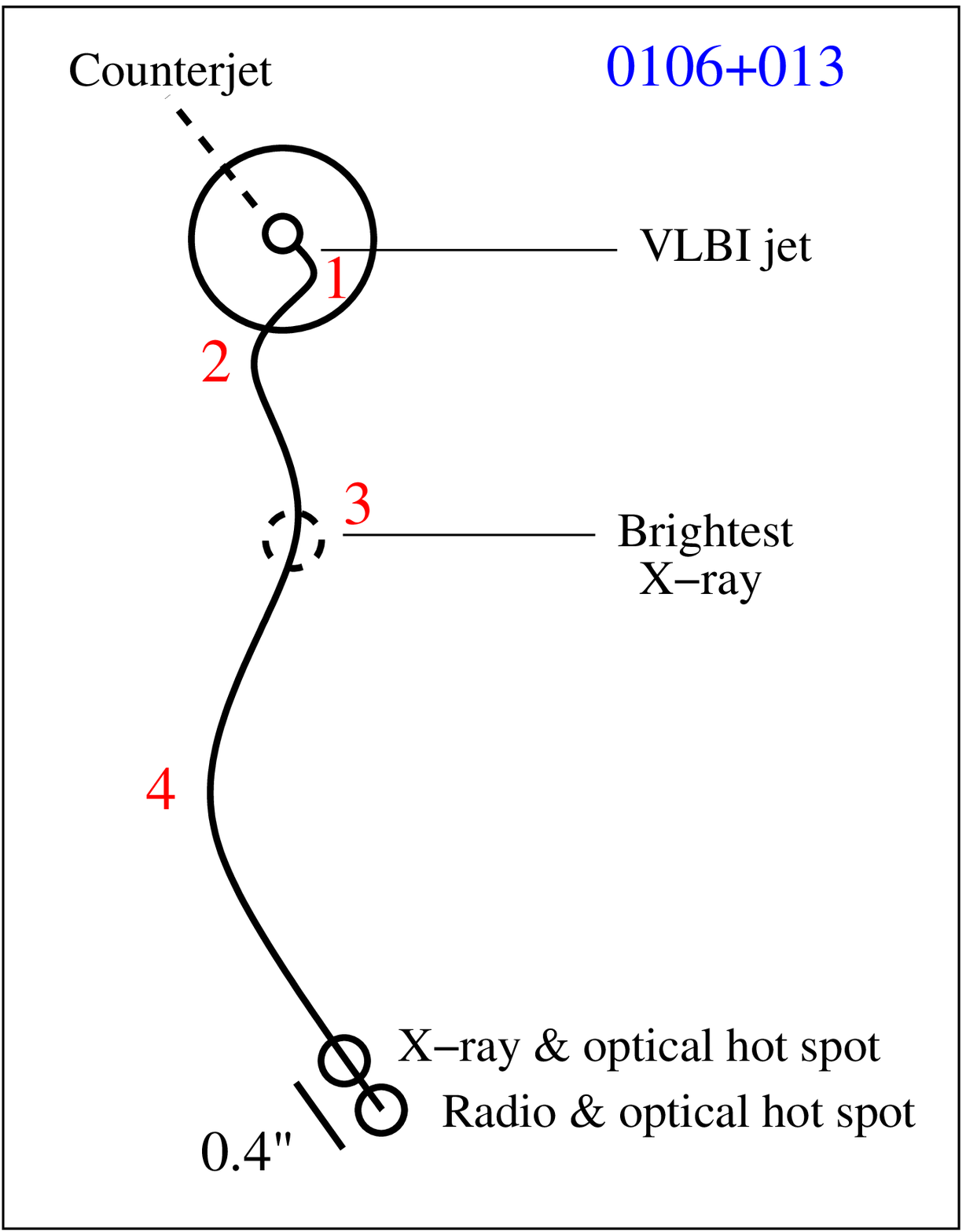}
\includegraphics[width=8cm]{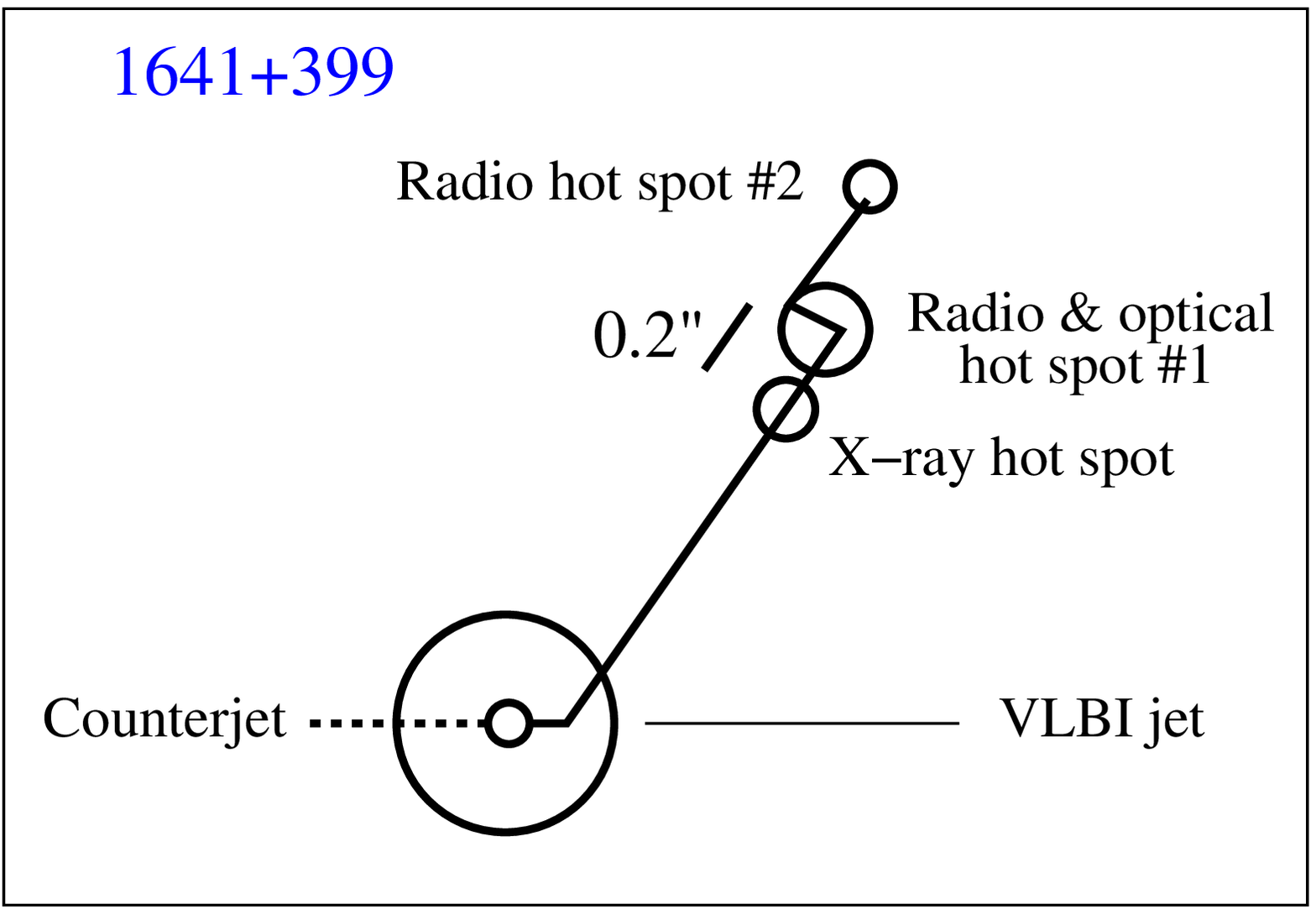}}
\caption{\small Schematics showing the kpc-scale jets in 0106+013 (left panel) and 3C\,345 (right panel).}
\label{figxfig0106}
\end{figure}

\section{Results}
\label{secspectral}
\subsection{Image Analysis: 0106+013}
The one-sided kpc-scale jet in 0106+013 suggests that the jet is relativistic up to hundreds of kiloparsecs. The X-ray jet is nearly as long as the radio jet (Figure \ref{figxrayimage}). It follows the radio jet trajectory, which has a gentle S-shape, extremely well. The X-ray jet first brightens (and is the brightest) at the point where the kpc-scale radio jet first changes trajectory and also becomes brighter. At the next radio bend, the X-ray jet emission is concentrated closer to one (outer) edge of the jet. The X-ray emission brightens once again just before the radio hot spot, and terminates before the radio hot spot. The surface brightness profiles of the radio and X-ray jets are displayed in Figures \ref{figradprofile} and \ref{figxrayprofile}. From these it is clear that the radio and X-ray peaks are offset by about 0$\farcs$4 ($\sim3.4$ kpc) in 0106+013. 

If the angle between the kpc-scale radio jet and line of sight is $\sim5\degr$ ({cf.} Section 4.4), then the deprojected jet length of 0106+013 is 38 kpc/sin(5$\degr$) = 436 kpc. As the parsec-scale jet has a position angle of $238\degr$, this implies an apparent pc-to-kpc-scale jet misalignment angle of $\sim54\degr$ \citep{Kharb10}. If the parsec-scale jet bends only once before 12 kpc or so (corresponding to the {\it Chandra} PSF of $\sim1\farcs$5) before aligning with the kpc-scale jet, then the radio jet has about four bends before it terminates at the hot spot. Figure \ref{figxfig0106} presents a schematic of the jet structure. The fact that the brightest X-ray emission is from bend\#3 is consistent with the idea that the Doppler factor is larger here as the jet inclination is smaller ($i.e.,$ the jet moves closer to our line of sight here). The jet plasma could therefore be following a helical path around the surface of a cone with a constant opening angle, the projected version of which is seen in Figures \ref{figxrayimage} - \ref{figxfig0106}. Two optical knots close to the jet terminal point are clearly detected in the {HST} image of 0106+013 (Figure \ref{fighst}). The knot closer to the core is coincident with the terminal point of the X-ray jet, while the knot further downstream is coincident with the radio hot spot. 

\subsection{Image Analysis: 3C\,345}
If the angle between the kpc-scale radio jet in 3C\,345 and the line of sight is $\sim5\degr$ ({cf.} Section 4.4), then the de-projected jet length is 25~kpc/sin(5$\degr$) = 287 kpc. The {\it Chandra} image of 3C\,345 reveals a short X-ray jet with a hot spot. The termination peak of the X-ray jet lies upstream of the radio hot spot (Figure \ref{fighst}). From the surface brightness profiles of the radio and X-ray jets displayed in Figures \ref{figradprofile} and \ref{figxrayprofile}, it is clear that the radio and X-ray peaks are offset by about 0.2$\arcsec$ ($\sim1.3$ kpc) in 3C\,345. 

The HST image of 3C\,345  image shows that the optical hot spot is not compact but 
{arc-like and extended, transverse} to the jet direction (Figure \ref{fighst}). The 4.9~GHz radio image also reveals a fainter, more compact hot spot further downstream from this optical hot spot. It appears as if the jet changed direction in this source by the angular extent of the optical ``hot spot'' (see Figure \ref{figxfig0106} for a schematic). This could be considered as being consistent with the suggestion that this source contains a binary black hole \citep[e.g.,][]{Lobanov05}, which could give rise to jet re-alignment on short timescales. The enhanced radio and optical emission from this jet bend could arise because the jet aligns closer to our line of sight in this section. There is no optical or X-ray emission in the region of the weak radio extension downstream of the bright hot spot. 

\begin{deluxetable}{lllllll}
\rotate
\tabletypesize{\scriptsize}
\tablecaption{Spectral Analysis Results}
\tablewidth{0pt}
\tablehead{\colhead{Source} &
\colhead{Component} & \colhead{Constraint} &\colhead{$N_{H}$ (Error)} & \colhead{$\Gamma$ (Error)} & \colhead{$F_{x}$ (Error)} & \colhead{$\chi^{2}/d.o.f.$}  \\
\colhead{} &\colhead{} & \colhead{} & \colhead{($\times10^{22}$\,cm$^{-2}$)} & \colhead{} & \colhead{($\times10^{-14}$\,erg\,cm$^{-2}$\,s$^{-1}$)} & \colhead{} }
\startdata
0106+013&Core (bin=50)   & $N_H$ free &3.6E$-$02  ($-$0.014, 0.012)          & 1.6 ($-$0.09, 0.04) & 125 ($-$3, 2)    & 166.42/149\\
&Jet\_1 (bin=20) & $N_H$ free & 1.6E$-$05  ($-$2.1E$-$06, 0.093) & 1.7 ($-$0.28, 0.43) & 2.9 ($-$0.1, 0.4)& 6.47/8\\
&                             & $N_H$ frozen & 0.028                                               & 1.8 ($-$0.30, 0.32) & 2.9 ($-$0.2, 0.2)& 7.00/9\\
&Jet\_2 (bin=15) &$N_H$ free & 2.2E$-$13 ($-$8.8E$-$14, 0.31)      & 2.1 ($-$0.49, 0.81) & 1.2 ($-$0.1, 0.5)& 4.37/3\\
&                            & $N_H$ frozen & 0.028                                                & 2.1 ($-$0.50, 0.63) & 1.2 ($-$0.1, 0.1)& 4.55/4\\
&Jet\_3 (bin=15) & $N_H$ frozen & 0.028                                               & 1.9 ($-$0.62, 0.78) & 1.1 ($-$0.1, 0.1)& 4.58/3\\  
1641+399&Core (bin=50)   & $N_H$ free& 3.0E$-$02 ($-$0.007, 0.005)            & 1.8 ($-$0.05, 0.05) & 384 ($-$5, 10) & 338.19/329\\
&Jet~~~(bin=20) & $N_H$ free& 0.14 ($-$0.14, 0.23)                         & 2.1 ($-$0.60, 0.93) & 2.5 ($-$0.5, 1)  & 7.58/8\\
&                             & $N_H$ frozen& 0.0104                                            & 1.8 ($-$0.30, 0.33) & 2.2 ($-$0.1, 0.1)& 8.75/9\\
\enddata
\tablecomments{
Results from fitting an absorbed powerlaw (XSPEC model = {\it phabs}({\it powerlaw})) to the X-ray data. For some fits, $N_{H}$ was frozen to the Galactic value which is 2.8$\times10^{20}$\,cm$^{-2}$ toward 0106+013 and 1.04$\times10^{20}$\,cm$^{-2}$ toward 3C\,345. Col.\,2: Source components (see Figure \ref{figregions}). Cols.\,4 and 5: Best-fit hydrogen column density and photon index, respectively, with errors quoted at the 90\% confidence level. Col.\,6: Unabsorbed 0.5$-$7.0 keV flux density with errors. {These errors were derived from the 90\% confidence level errors in the ``norm'' parameter derived in the model-fits.} Col.\,7: $\chi^{2}$ by degrees of freedom.}
\label{tabxspec}
\end{deluxetable}

\begin{figure}
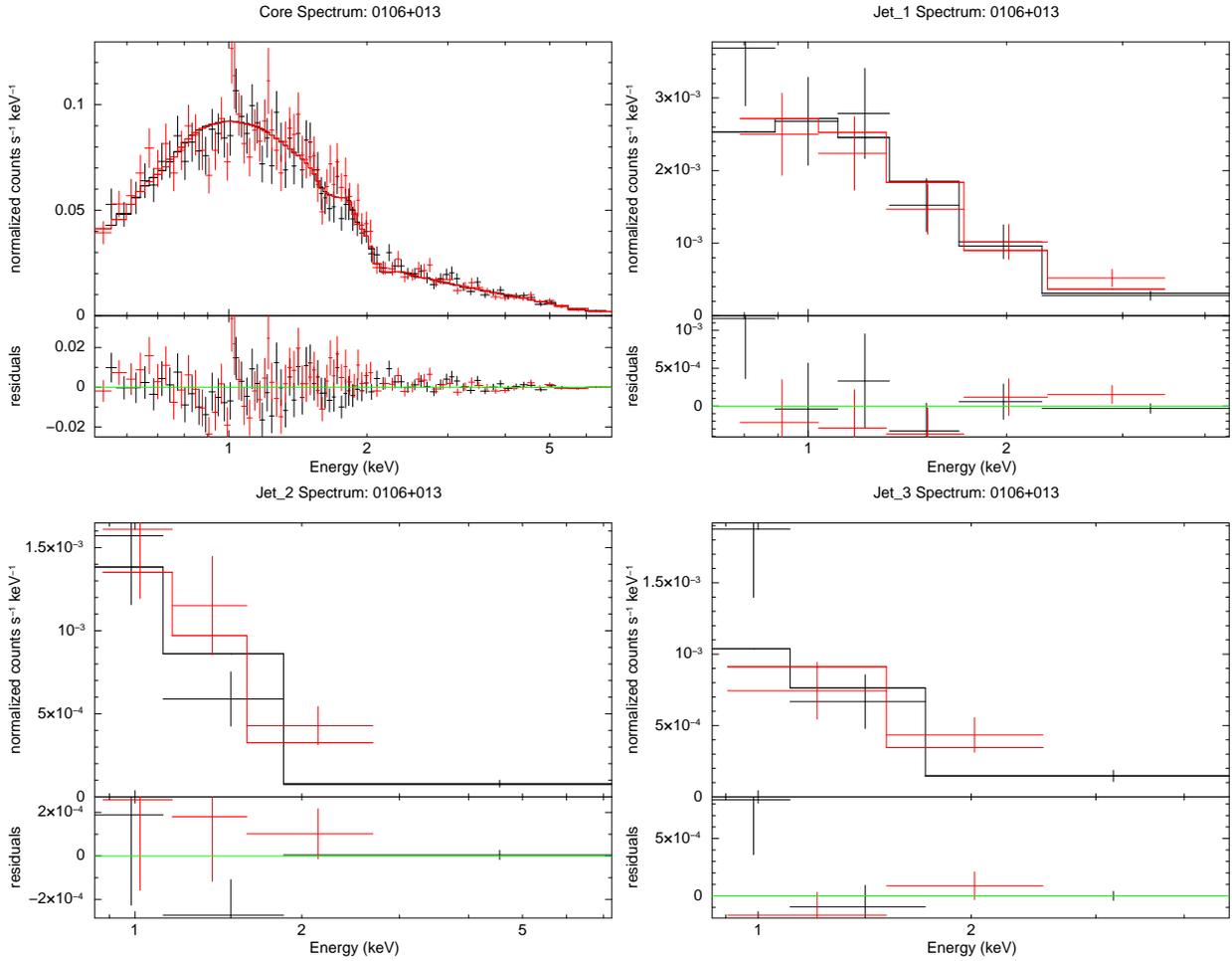

\centering{
\includegraphics[width=6.4cm,angle=270]{f7a.ps}
\includegraphics[width=6.4cm,angle=270]{f7b.ps}
\includegraphics[width=6.4cm,angle=270]{f7c.ps}
\includegraphics[width=6.4cm,angle=270]{f7d.ps}}
\caption{\small Spectral analysis results for an absorbed powerlaw model for the core (top left), ``jet\_1" (top right), ``jet\_2" (bottom left) and ``jet\_3" component (bottom right)  of 0106+013. The residuals of the model-fit are displayed at the bottom of each panel. The red and black data denote the two individual exposures, which were fit simultaneously.}
\label{figspectral0106}
\end{figure}
\begin{figure}
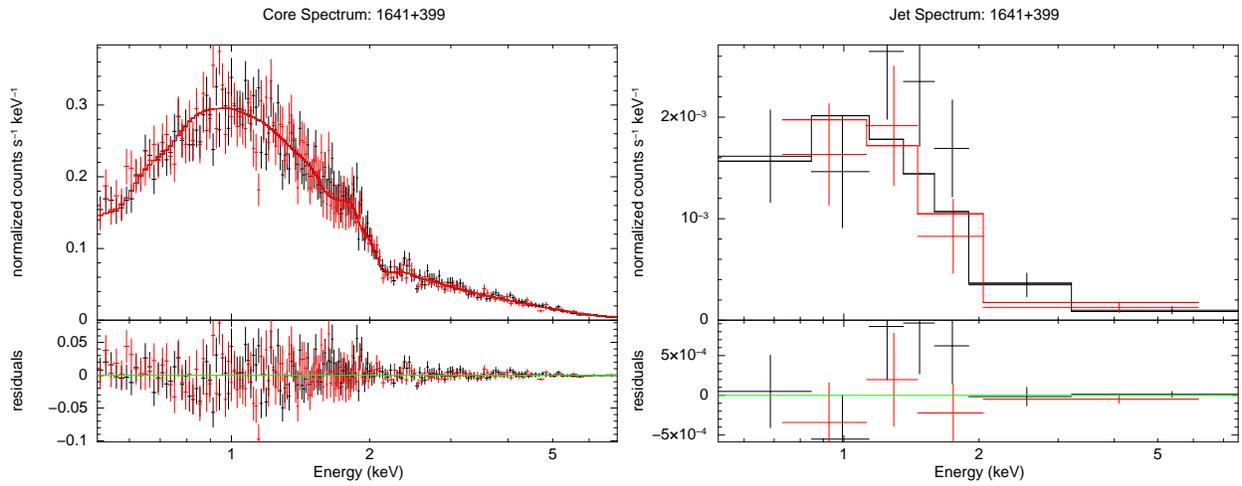

\centering{
\includegraphics[width=6.4cm,angle=270]{f8a.ps}
\includegraphics[width=6.4cm,angle=270]{f8b.ps}}
\caption{\small Spectral analysis results for an absorbed powerlaw model for the core (left panel) and jet component/hot spot (right panel) of 3C\,345. The residuals of the model-fit are displayed at the bottom of each panel. The red and black data denote the two individual exposures, which were fit simultaneously.}
\label{figspectral1641}
\end{figure}

\subsection{X-ray Spectral Analysis: 0106+013 \& 3C\,345}
\label{secxspec}
The core and jet/hot spot emission for both 0106+013 and 3C\,345 were fitted with a powerlaw model with photoelectric absorption (XSPEC model = {\it phabs}({\it powerlaw})), where we allowed the hydrogen column density ($N_{H}$) to first, be freely variable, and second, to be frozen to the Galactic $N_{H}$ value. We found that the best-fit $N_{H}$ values were not well constrained for the jet or hot spot components of either 0106+013 or 3C\,345. We therefore regard the estimates obtained with $N_{H}$ fixed to the Galactic value as more robust. The small number of X-ray counts (typically $<$100) precluded an analysis with more sophisticated spectral models. The results of the spectral model-fitting, along with the error estimates at the 90\% confidence level, are presented in Table \ref{tabxspec}, while the best-fit core and jet models along with the residuals are displayed in Figures \ref{figspectral0106} to \ref{figspectral1641}. The best-fit jet models have the $N_{H}$ value frozen to the Galactic value toward the source direction. The 1 keV flux estimates are listed in Table \ref{tabmulti}.

\begin{deluxetable}{ccccccc}
\tablecaption{Flux Estimates from the Multiwavelength Data}
\tablewidth{0pt}
\tablehead{\colhead{Source}&\colhead{Comp}&\colhead{$S_{r}^{1.6}$}&\colhead{$S_{r}^{4.9}$}&\colhead{$S_{o}^{4747}$} & \colhead{$S_{o}^{5850}$} &\colhead{$S_{x}^{1\,keV}$}}
\startdata
0106+013&Core   & 5.2E$-$14 & 1.2E$-$13 & ... & 7.4E$-$13 & 6.8E$-$15 \\
                   &Jet\_1 & 3.7E$-$16 & 3.8E$-$16 & ... & 2.0E$-$16 & 1.9E$-$16 \\
                   &Jet\_2 & 6.4E$-$16 & 6.5E$-$16 & ... & 2.4E$-$16 & 9.9E$-$17 \\
                   &Jet\_3 & 4.9E$-$15 & 4.8E$-$15 & ... & 1.5E$-$16 & 7.5E$-$17 \\
1641+399&Core   & 1.2E$-$13 & 4.1E$-$13 & 3.7E$-$12 & 1.5E$-$12 & 2.2E$-$14 \\
                   &Jet      & 6.0E$-$15 & 8.9E$-$15 & 2.0E$-$15 & 2.0E$-$15 & 1.4E$-$16 
\enddata
\tablecomments{\small Cols.\,1 \& 2: Source names and components, respectively (see Figure \ref{figregions}). Cols.\,3 to 7: Flux estimates in ergs\,cm$^{-2}$\,sec$^{-1}$ at 1.6 GHz (VLA), 4.9 GHz (VLA), 4747 \AA~(HST/ACS), 5850 \AA~(HST/STIS), and 1 keV (Chandra), respectively.}
\label{tabmulti}
\end{deluxetable}

\subsection{Spectral Energy Distributions and Modeling}
\label{secsed}
\begin{figure}
\centering{
\includegraphics[width=8.1cm]{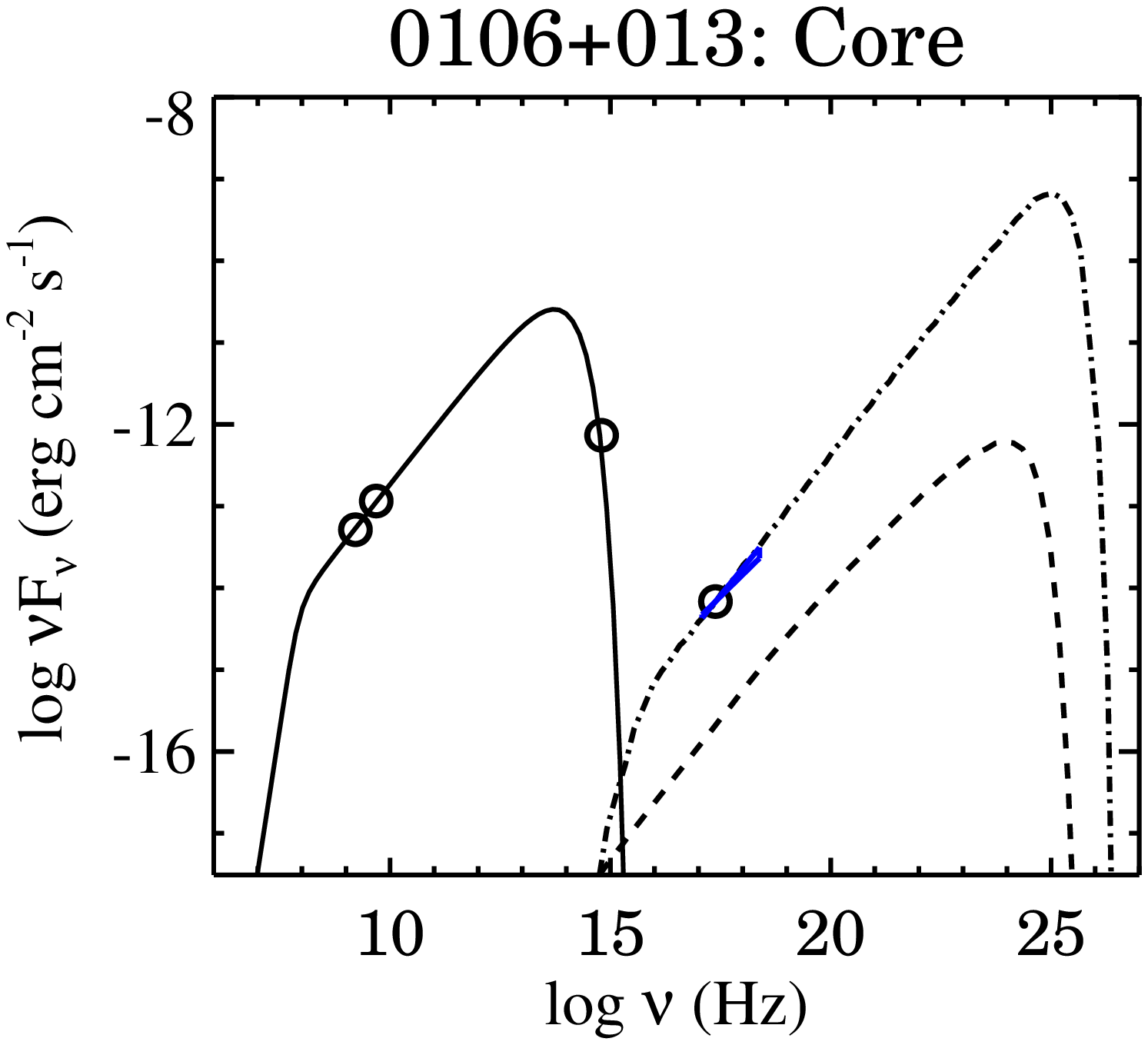}
\includegraphics[width=8.1cm]{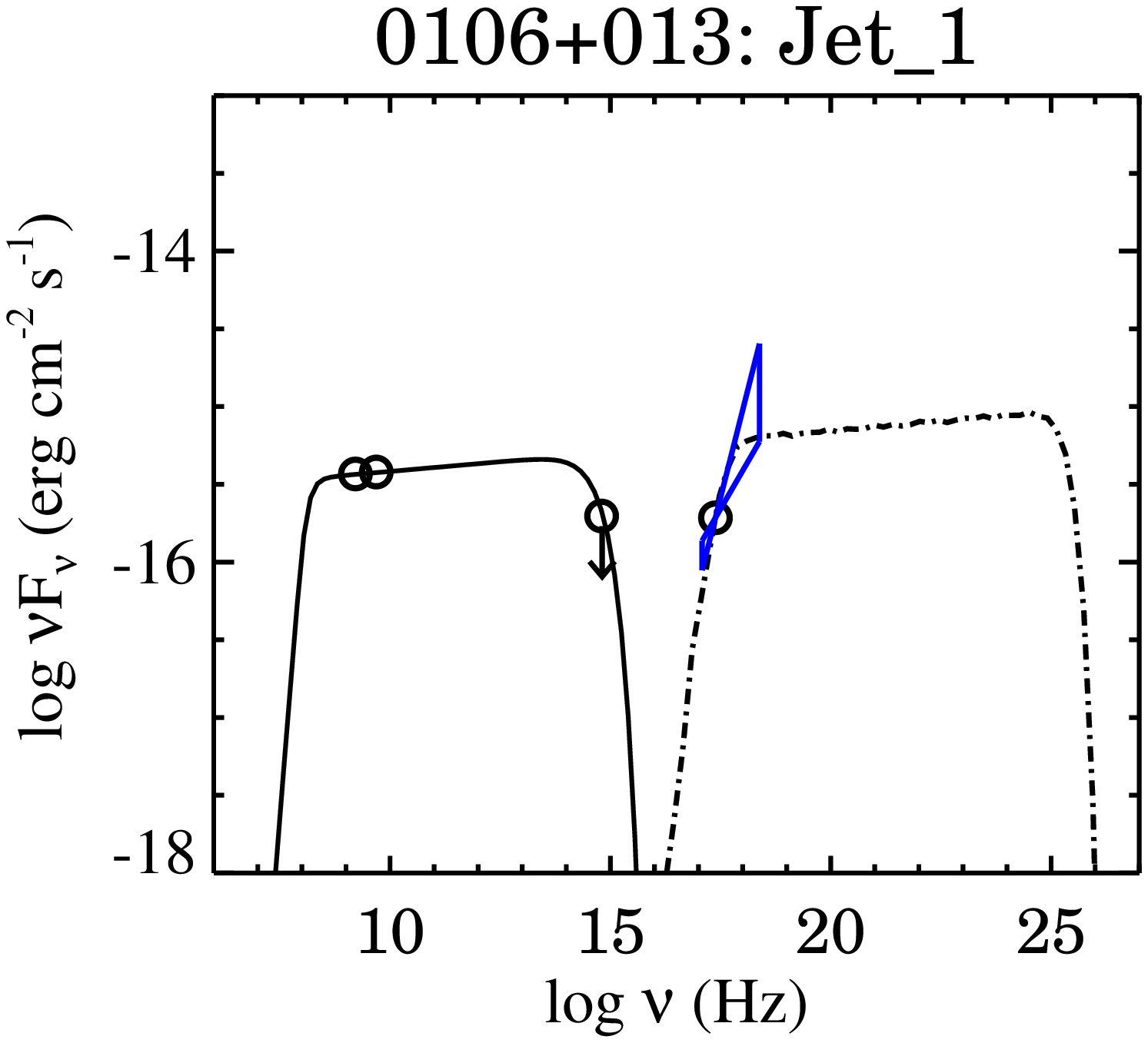}
\includegraphics[width=8.1cm]{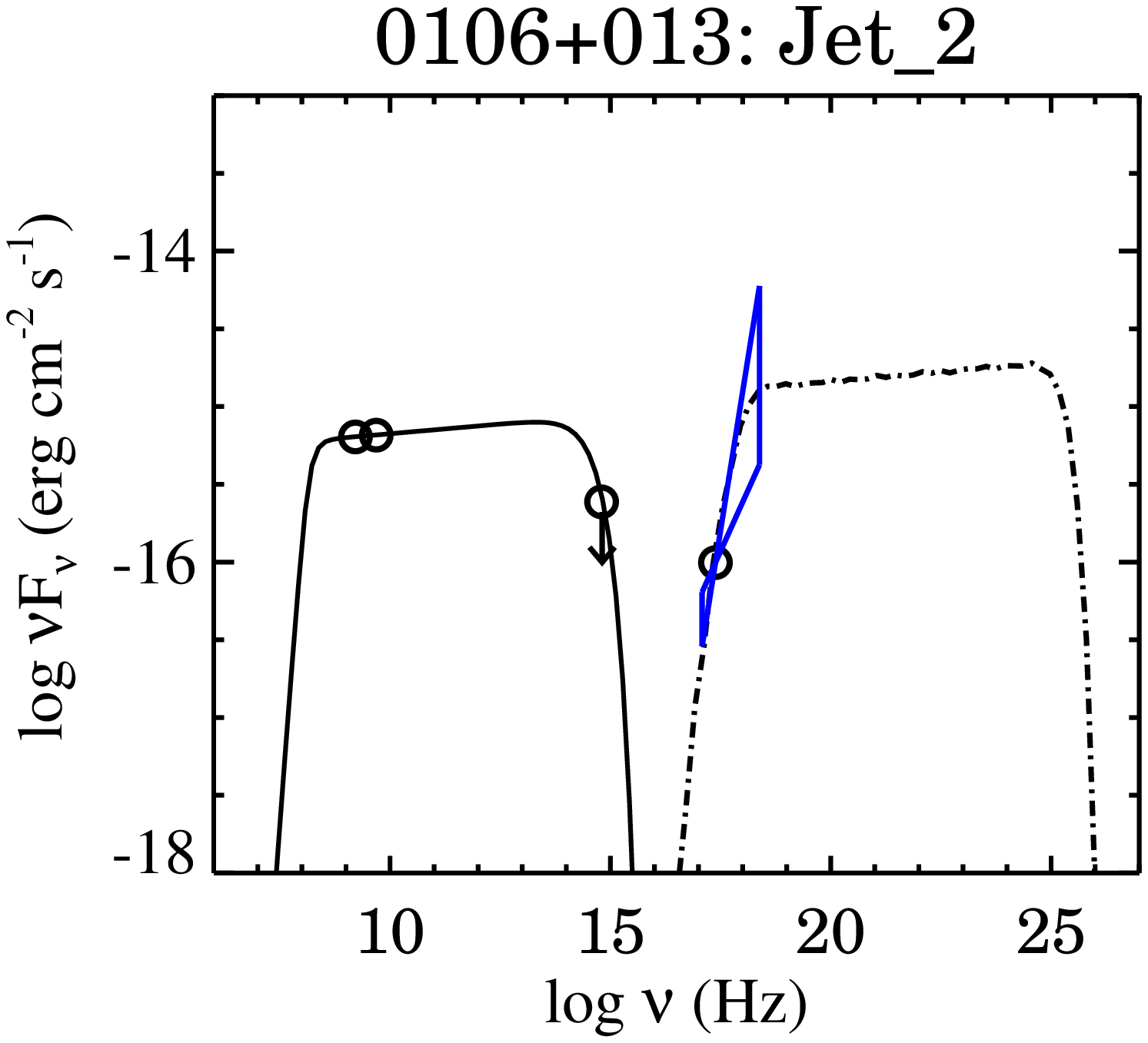}
\includegraphics[width=8.1cm]{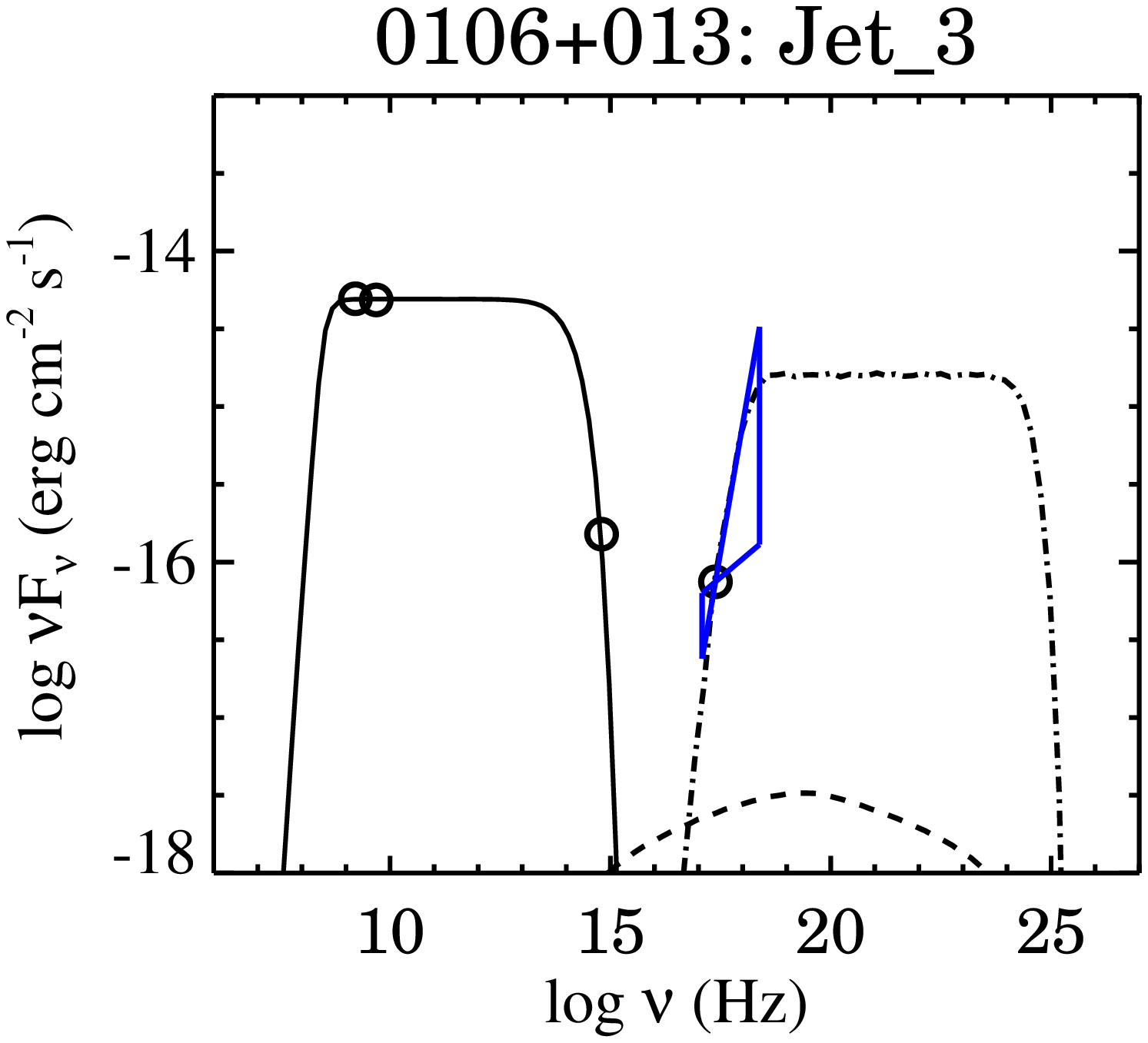}}
\caption{\small Broad-band SED model-fits for the 0106+013 core (top left), ``jet\_1" (top right),
``jet\_2" (bottom left) and ``jet\_3" (bottom right). The solid line, dot-dashed and dashed lines represent the synchrotron, IC and SSC models, respectively, while the data are represented by the open circle symbols. The symbol sizes are indicative of the flux error estimates. The best-fit parameters are listed in Table \ref{tabsed}. }
\label{figsed0106core}
\end{figure}

\begin{figure}
\centering{
\includegraphics[width=8.1cm]{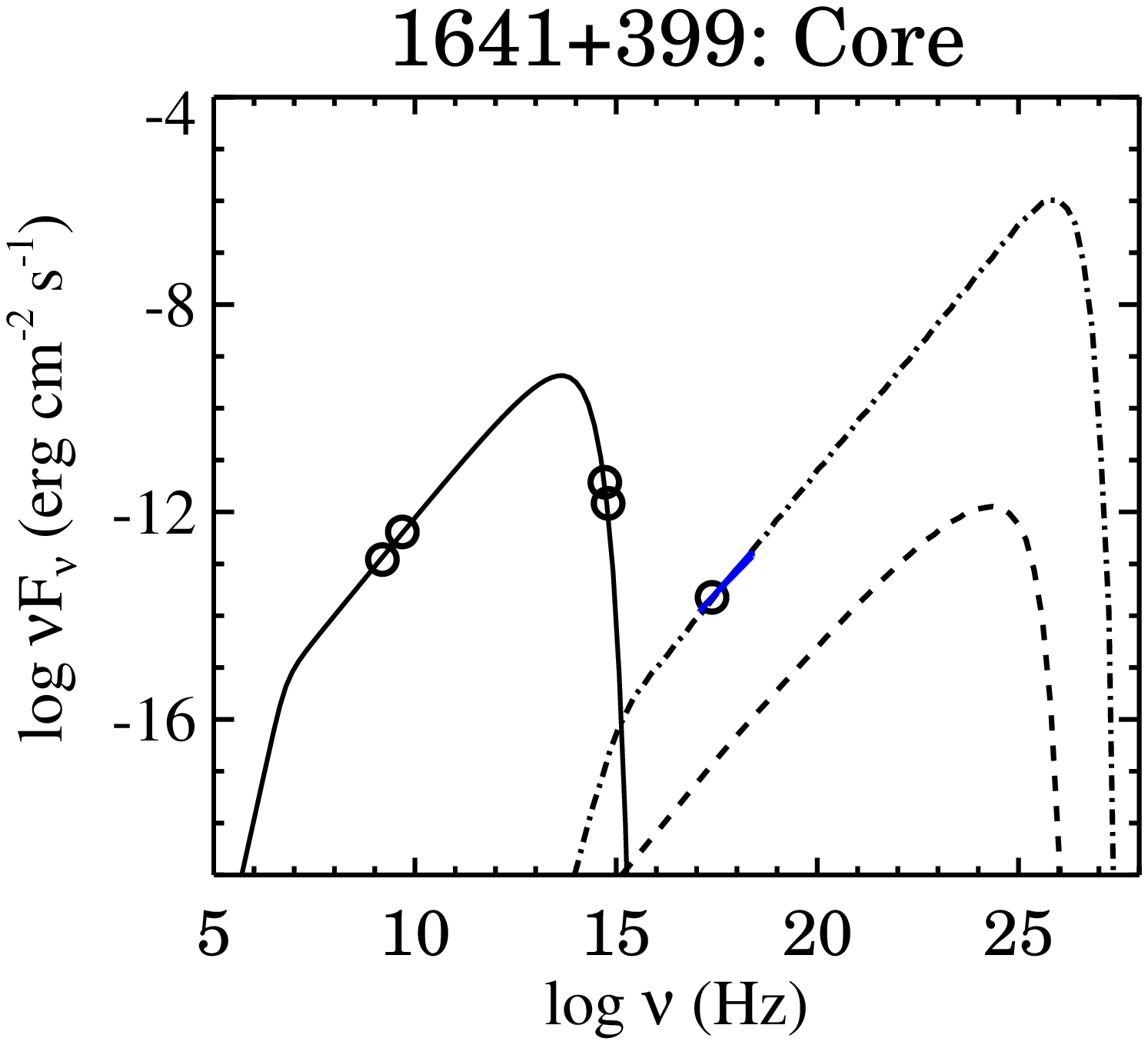}
\includegraphics[width=8.1cm]{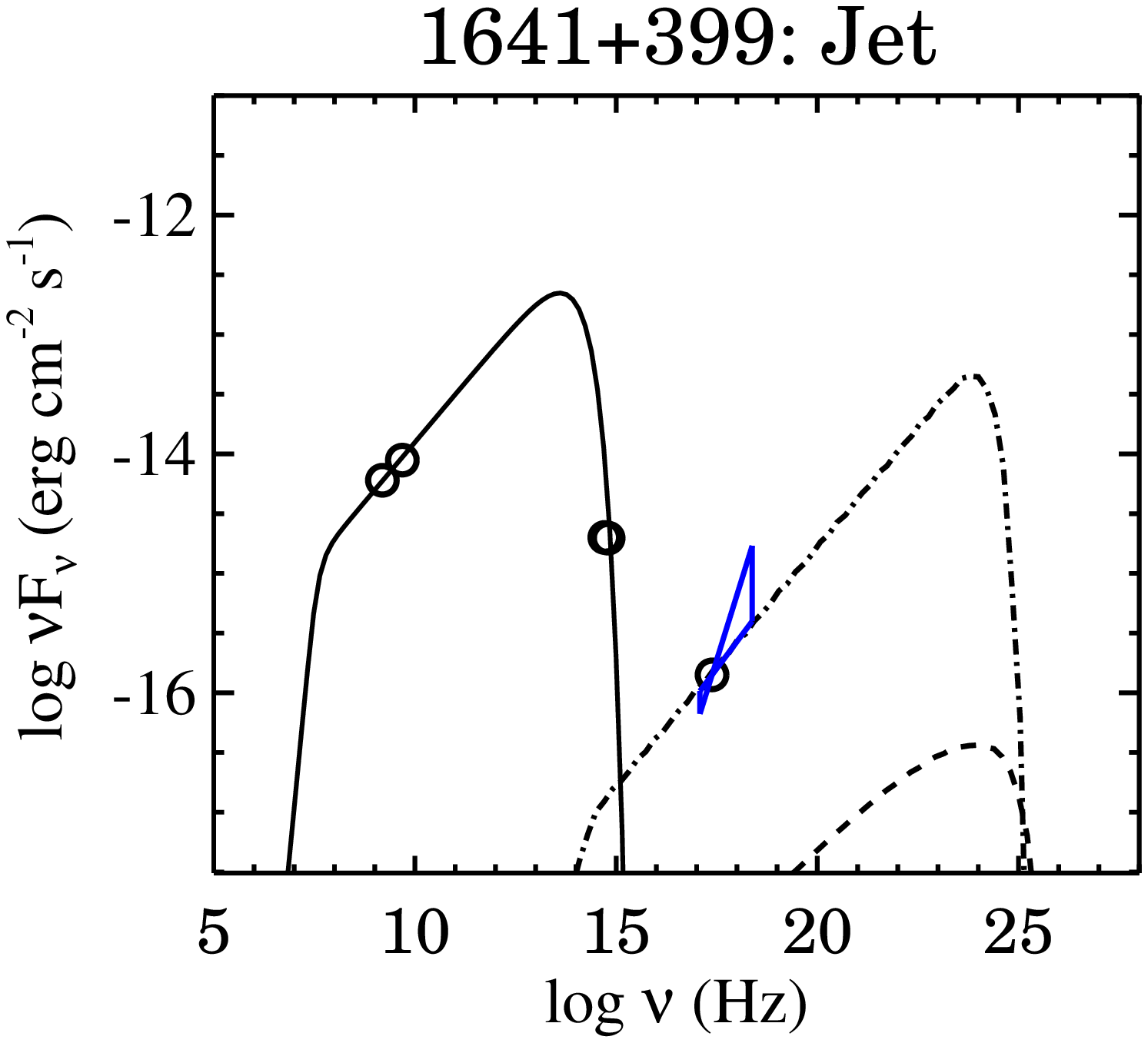}}
\caption{\small Broad-band SED model-fits for the 1641+399 core (left panel) and hot spot (right panel). The solid line, dot-dashed and dashed lines represent the synchrotron, IC and SSC models, respectively, while the data are represented by open circles. The best-fit parameters are listed in Table \ref{tabsed}. }
\label{figsed1641jet}
\end{figure}

We constructed broad-band (radio-optical-X-ray) SEDs for the core and jet components of the two sources, and tried model-fitting with the synchrotron, IC and SSC emission models, using the {\it modified} SED code of \citet{Xue08}. The original SED code\footnote{http://jelley.wustl.edu/multiwave/spectrum/download.htm} from \citet{Krawczynski04} was modified by Xue et al., to include IC scattering off CMB photons. A single powerlaw was assumed for the electron population. The broad-band SED model-fitting results for the core and jet components of 0106+013 and 3C\,345 are tabulated in Table \ref{tabsed}, and displayed in Figures \ref{figsed0106core} to \ref{figsed1641jet}. For the creation of the 3C\,345 jet/hot spot SED, the small ($\sim0\farcs2$) offset between the radio, optical and X-ray peak emission, was ignored. The errors associated with fitting the electron powerlaw indices are of the order of 0.05. Similarly, the errors in fitting $\gamma_{min}$ and $\gamma_{max}$ values are of the order of 10 to 15, and 1E+5 to 1.5E+5, respectively. The Doppler factors are constrained by the X-ray bow-tie slopes to within a few. In turn, the magnetic field strengths are restricted to within a few to few tens of $\mu$G.

The SED modeling primarily suggests that the optical emission has a synchrotron origin in both the jets, while the X-ray emission is primarily from the IC/CMB mechanism throughout. We note that even though the X-ray data lies roughly along or below the extrapolation of the radio-optical slope in the $S_{\nu}$ vs. $\nu$ plots in all instances, the X-ray slopes (as represented by the bow-ties in Figures \ref{figsed0106core} $-$ \ref{figsed1641jet}), preclude the synchrotron model as the primary emission mechanism for the X-rays. The SSC model component is also plotted in the SED model-fits. In all cases, the SSC model predictions lie well below the observed data. We discuss the implications of the best-fit parameters in Section \ref{secdiscussion}.

\begin{deluxetable}{lll}
\tablecaption{SED best fit parameters}
\tablewidth{0pt}
\tabletypesize{\small}
\tablehead{\colhead{Source}&\colhead{Feature}&\colhead{Fitted Parameters}}
\startdata
0106+013 & Core & $\delta$ = 11, B = 60 $\mu$G,~~~~$\gamma_{min}$ = 4, ~~~~\,$\gamma_{max}$ = 2E+5, ~n = 1.63 \\
                    & Jet\_1 & $\delta$ = 3,~~\,B = 54 $\mu$G,~~~~$\gamma_{min}$ = 125, ~\,$\gamma_{max}$ = 1E+6, ~n = 2.95 \\
                    & Jet\_2 & $\delta$ = 3,~~\,B = 53 $\mu$G,~~~~$\gamma_{min}$ = 175, ~\,$\gamma_{max}$ = 7E+5, ~n = 2.95 \\
                    & Jet\_3 & $\delta$ = 3,~~\,B = 140 $\mu$G,~~$\gamma_{min}$ = 211, ~\,$\gamma_{max}$ = 3E+5, ~n = 3.00\\
1641+399 & Core & $\delta$ = 45,~B = 7 $\mu$G,~~~~~\,$\gamma_{min}$ = 1, ~~~~\,$\gamma_{max}$ = 2E+5, ~n = 1.10 \\
                    & Jet & $\delta$ = 5,~~\,B = 75 $\mu$G,~~~~$\gamma_{min}$ = 3, ~~~~\,$\gamma_{max}$ = 2E+5, ~n = 2.20 \\
\enddata
\tablecomments{\small $\delta$ = Doppler factor, B = magnetic field strength, $\gamma_{min}$, $\gamma_{max}$ = minimum and maximum electron Lorentz factors, respectively, n = electron powerlaw index. {The radius of the emission volumes was 2.5E+22 cm and 2.0E+22 cm for 0106+013 and 1641+399, respectively.} 
The CMB photon energy density was assumed to be 3.7E$-$11 erg~cm$^{-3}$ for 0106+013, and 2.6E$-$12 erg~cm$^{-3}$ for 1641+399. The errors in the fitted parameters are discussed in Section \ref{secsed}.}
\label{tabsed}
\end{deluxetable}

\begin{figure}
\centering{
\includegraphics[width=8.1cm]{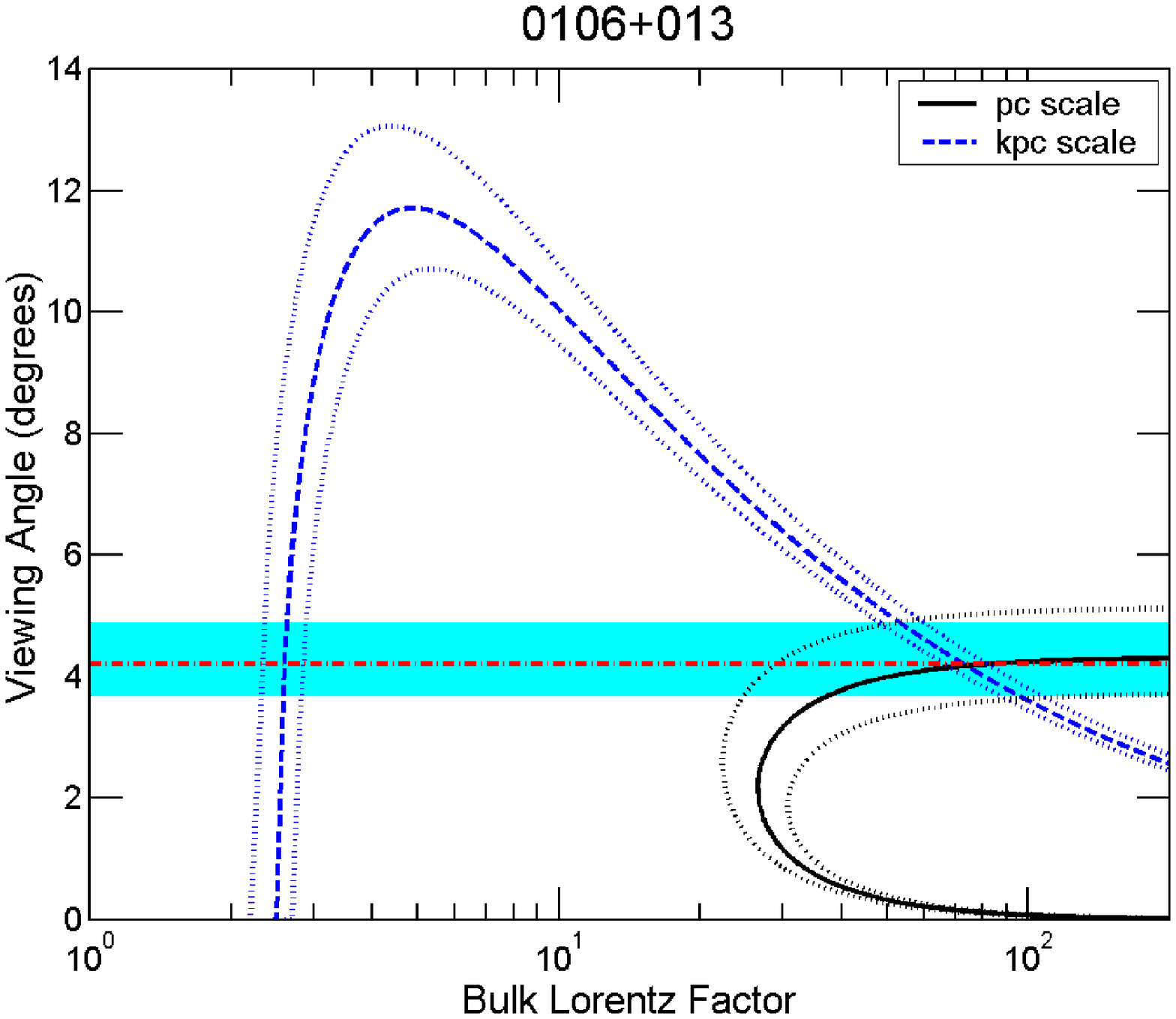}
\includegraphics[width=8.1cm]{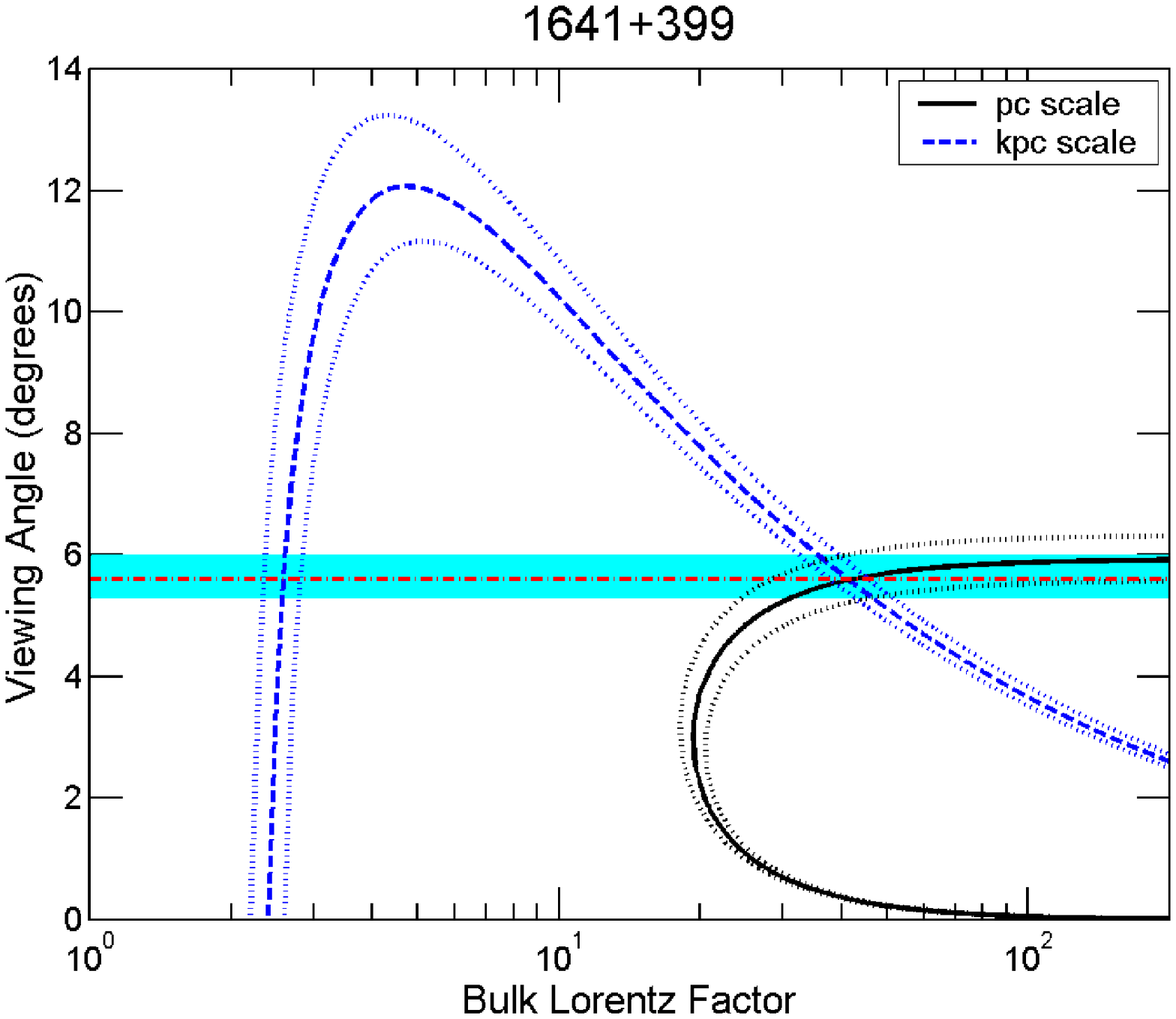}}
\caption{\small Plot of viewing angle versus the bulk Lorentz factor of the jet. The black curve (parsec scale) is defined by equation 1, while the blue curve (kpc scale) is defined by equation 3 ({cf.} Section \ref{secK}). The intersection of the two curves indicates the Lorentz factor and viewing angle of the jet, in the case of no bending or deceleration between pc- and kpc-scales. The cyan shaded region represents the possible range of jet inclination angles if the jet decelerates from pc- to kpc-scales, but does not bend.}
\label{figK}
\end{figure}

\section{Constraints on Jet Speed and Inclination from Pc- and Kpc-scale Data}
\label{secK}
\citet{Hogan11} found that the extended radio flux densities were closely correlated with the detection rate of the X-ray jets in the MCS: the jet detection was 100\% for jets with 1.4 GHz extended flux density $>$300 mJy, but only $\sim60$\% for jets below this value. As \citet{Kharb10} have reported a significant trend between the parsec-scale apparent jet speeds and the extended radio luminosity in MOJAVE blazars, it was suggested by Hogan et al. that there could be a link between the X-ray jet detection and jet speed. This suggestion is fully consistent with the expectations of the IC/CMB beaming model, since faster jets would have an increased CMB photon density (by a factor $\Gamma_{jet}^2$) in the jet rest-frame. Therefore, through the use of the parsec-scale apparent jet speed data, a connection to the kpc-scale X-ray emission was indicated from the {\it Chandra} pilot study. Furthermore, model-dependent constraints on the jet inclination and Lorentz factors were derived by \citet{Hogan11}, using parsec and kpc-scale radio and X-ray data. 
Following the same approach, we present below the results from our newer, more sensitive data.

Assuming that the jet speed, $\beta$ ($\equiv v/c$), and jet inclination, $\mu$ (= cos\,$\theta$), remains constant going from parsec to kiloparsec scales, the jet speed can be expressed as a function of the parsec-scale apparent jet speed, $\beta_{app}$, and the angle to the line of sight, $\theta$, as,
\begin{equation}
\beta=\frac{\beta_{app}}{\beta_{app}\mu+\sqrt {1-\mu^{2}}}.
\end{equation} 
\citet{Marshall05} defined the K parameter as a combination of the magnetic field strength and the ratio of the observed X-ray (assumed to be inverse Compton) to radio (synchrotron) luminosities of the kpc jets through the relation,
\begin{equation}
K=B_{eq} (aR)^{1/(\alpha_{r}+1)} (1+z)^{-(\alpha_{r}+3)/(\alpha_{r}+1)} b^{(1-\alpha_{r})/(\alpha_{r}+1)}
\end{equation}
where $a=9.947\times10^{10}$ G$^{-2}$, $b=3.808\times10^{4}$ G \citep[e.g.,][]{HarrisKrawczynski02} and $\alpha_{r}$ is the radio spectral index (= 0.8). The $a$ and $b$ parameters were found by equating the expected and observed values of the ratio of X-ray to radio energy densities, $R$ (see Table \ref{tabderived}), under the equipartition assumption. For a jet with Lorentz factor $\Gamma_{jet}>1.5$, K can be defined as a function of jet speed and inclination angle as,
\begin{equation}
K=\frac{1-\beta+\mu-\beta\mu}{(1-\beta\mu)^{2}}.
\end{equation}

The relationship between the K parameter and the jet speed $\beta$ can thus be explored for any given source. Such an attempt has been made in Figure \ref{figK} for the two quasars. The black curve (parsec scale) is defined by equation 1, while the blue curve (kpc scale) is defined by equation 3. The intersection of the blue dashed and black solid curves in Figure \ref{figK} provides a single ($\mu,\Gamma_{jet}$) pair that satisfies equations 1 and 3. 

If we relax the assumption of constant Lorentz factor and allow for the possibility of deceleration between parsec and kpc scales, then the kpc-scale bulk Lorentz factor is given by the intersection of the red dashed line with the tail of the blue K curve at low $\Gamma_{jet}$. The cyan shaded region represents the possible range of jet inclination angles if the jet decelerates from parsec to kpc scales, but does not bend. We note that the observed K values indicate upper limits of $\sim13\degr$ on the jet inclination angle for the two quasars, irrespective of any assumptions about bending or deceleration. The errors in the K parameter (in Table \ref{tabderived} and Figure \ref{figK}) have been estimated using a Monte Carlo error analysis approach described by \citet{Hogan11}.

This method can yield large or even extreme $\Gamma_{jet}$ values in some jets. 0106+013 is specifically one of the jets in the MCS which has a large bulk Lorentz factor when examined in this way ($\Gamma_{jet}\sim70$). 3C\,345 has a $\Gamma_{jet}$ value of $\sim$40, which places it in the uppermost range of observed speeds for the bright radio blazar population \citep{Lister09b}, although still not as high as 0106+013. If we relax the assumption of constant $\Gamma_{jet}$ between parsec and kiloparsec scales, then the best fit value of the kpc-scale $\Gamma_{jet}$ becomes $\sim2.5$. It is likely that both jet deceleration and jet bending may need to be invoked in these jets for a more realistic picture \citep[see][]{Hogan11}.
  
\section{Summary and Discussion}
\label{secdiscussion}
{
\subsection{PKS B0106+013}
The one-sided kiloparsec jet of 0106+013 displays distinct jet bends, which appear to increase in amplitude further along the jet. There is a distinct lack of radio lobes in this source. Instead, the radio jet morphology resembles that of a ``nose-cone'' \citep[e.g.,][]{Clarke86}. Magnetized plasma appears to move forward through the terminal shocks in the jets, rather than backward like in radio lobes \citep[e.g.,][]{Komissarov99}.
The X-ray emission follows the one-sided radio jet closely. There is no X-ray emission from the counterjet direction. X-ray emission peaks at the site of the first prominent jet bend on kpc scales (bend \#3 in Figure \ref{figxfig0106}). The jet could be pointing closer to our line of sight at the position of this bend, which could be making both the radio and the X-ray emission brighter due to Doppler boosting. As there is no optical counterpart to the radio and X-ray peak emission at this jet bend, the X-ray emission is likely to be from the IC/CMB mechanism. If the jet bend was due to jet-medium interaction, deceleration in the jet flow should have occurred, which would have in turn decreased the X-ray intensity after the bend, instead of increasing it, as is observed. Therefore, we favor the picture of a kpc-scale jet with a ridge line that wraps around the surface of a cone, following a helical path.

Around the second prominent jet bend (bend \#4 in Figure \ref{figxfig0106}), the X-ray emission becomes more prominent at the outer edge of the radio jet. This ``edge-brightening'' could be a result of  particle acceleration in a ``sheath'', as proposed by \citet{Stawarz02}. It could also imply that the electrons in a slower moving sheath upscatter relativistically boosted photons from the faster-moving jet spine \citep{Ghisellini05}. As this portion of the jet is expected to be moving away from our line of sight in the helical jet picture, it could explain why the spine emission fails to dominate in the X-ray emission here. After the second jet bend, however, the bulk of the X-ray emission shifts to the central regions of the radio jet, as if arising in the jet ``spine''. Finally, the X-ray jet terminates about 0\farcs4 ($\sim3.4$ kpc) upstream of the radio hot spot. The X-ray termination peak could be indicative of the onset of bulk deceleration in the kpc scale jets. As the jet Lorentz factor decreases further downstream, the X-ray emission, which has an inverse Compton origin (as indicated by the X-ray slopes in the multiwaveband SEDs in Figure \ref{figsed0106core}), decreases due to the decrease in the relativistically boosted CMB photons seen in the rest frame of the jet. 

The {HST} image of 0106+013 source shows two clear jet knots close to its terminal point - one of them corresponds to the termination point of the X-ray jet, while the second one is coincident with the radio hot spot. The optical counterpart of the X-ray termination point has a synchrotron origin (see Figure \ref{figsed0106core}). We expect that the shocks that are associated with the bulk deceleration sites and the radio hot spots are likely to reaccelerate and re-energize particles that give rise to optical synchrotron emission. 
The presence of the two discrete optical hot spots suggests that the shocked regions are highly localized.

\subsection{3C\,345}
The kpc-scale radio image of 1641+399 or 3C\,345 exhibits a bright core and hot spots along the jet and counterjet direction, with the hot spot-core-counter-hot spot forming a V-shape \citep{Cooper07}. The X-ray jet of 3C\,345 follows the short projected radio jet up until its bright ``hot spot'' (hot spot \#1 in Figure \ref{figxfig0106}). There is no X-ray emission from the counterjet direction. The X-ray jet termination peak is found upstream of the radio hot spot by around 0$\farcs$2 ($\sim$1.3 kpc). While the radio jet continues after this hot spot, the X-ray jet does not. HST optical emission is seen in an arc-like structure, well coincident with the bright radio hot spot. We propose that this feature is a sharp jet bend rather than a termination peak. A weak radio hot spot (hot spot\#2 in Figure \ref{figxfig0106}) is indeed observed less than $1\arcsec$ downstream of the bright radio hot spot, but does not have an optical or X-ray counterpart. It is likely that the optical and radio emission is enhanced in the sharp jet bend as it approaches closer to our line of sight. On the other hand, the sharp jet bend could reflect rapid jet deflection in this binary black hole candidate.

\subsection{Constraints on Jet Speed \& Inclination Angle} 
Using the parsec-scale radio and kpc-scale radio-X-ray data, we have derived constraints on the jet inclination angles and bulk Lorentz factors for the two quasars. Assuming a constant jet speed from parsec to kpc scales, a $\Gamma_{jet}$ of $\sim$70 is derived for 0106+013, and $\sim$40 for 3C\,345. On relaxing this assumption, the best fit value of the kpc-scale Lorentz factor becomes $\sim$2.5. Upper limits of $\sim13\degr$ on the jet inclination angle are obtained for the two quasars, irrespective of any assumptions about jet bending or deceleration. It is likely that both jet deceleration and jet bending may need to be invoked in these jets for a more realistic picture \citep{Hogan11}.

\subsection{Radio-X-ray Offsets}
As the X-ray terminal peak emission is likely to be coincident with sites of bulk jet deceleration, the offsets between the X-ray terminal emission and the radio hot spots of a few kiloparsecs as observed in 0106+013 and 3C\,345, suggest that the radio hot spot is {\it not} the primary site of bulk jet deceleration in these jets \citep[e.g.,][]{Hardcastle07}. This suggestion gives credence to the idea that jet instabilities rather than jet interaction with the surrounding medium are the primary cause of the bulk jet deceleration. Offsets between the X-ray and radio hot spots have been observed in several jets. For example, in the jets of quasars 4C\,74.26 \citep{Erlund07}, PKS 1055+201 \citep{Schwartz06}, PKS 2101$-$490 (Godfrey et al., 2011, ApJ, submitted), and in the jets of radio galaxies 3C\,390.3, 3C\,227 \citep{Hardcastle07}, 3C\,353 \citep{Kataoka08} and 3C\,445 \citep{Perlman10}. 

\subsection{SED Modeling Trends}
Broad-band spectral energy distribution modeling of individual core/jet components in both 0106+013 and 3C\,345, suggests that the optical emission is from the synchrotron mechanism, while the X-rays are produced via the inverse Compton mechanism from relativistically boosted cosmic microwave background seed photons.
We see from the best-fit parameters in Table \ref{tabsed}, that there is a trend of increased magnetic field strength at the jet termination regions compared to the cores, for both the sources. This could be indicative of shocked regions that locally enhance the magnetic field strengths, as well as reaccelerate particles, close to the jet termination points. There is also a trend of a steeper electron powerlaw index in the jet regions as compared to the cores. While the $\gamma_{max}$ values do not show a clear trend between the various core/jet components, the $\gamma_{min}$ values seem to increase downstream. Overall, these results are consistent with shocks amplifying the magnetic fields and reaccelerating charged particles toward the terminal regions of the jets. In highly magnetized jets, these ``shocks'' could be sites of magnetic field dissipation instead, and could also reaccelerate charged particles.

\subsection{The Case for Magnetized Jets}
\citet{Nakamura07} have suggested that the wiggling structures in AGN jets could be a result of current-driven instabilities in highly magnetized jets. These instabilities are also capable of accelerating particles to high enough energies to emit X-ray and gamma ray photons \citep[e.g.,][]{Lyutikov03}. However, it has also been suggested that jets are likely to be Poynting flux dominated close to the jet launching sites, on scales of less than a parsec, but become particle dominated on kiloparsec scales \citep[e.g.,][]{Sikora05}. 
Whether AGN jets are highly magnetized or particle-dominated, and if/where in the jet such a transition occurs, is a highly contentious issue. TeV gamma ray flux variability has both been explained by particle-dominated jets via broadband SED modeling \citep[e.g., in the blazar Mrk\,421,][]{Acciari11}, and by the ``jet-in-jet'' model inside Poynting flux dominated jets \citep[e.g.,][]{Giannios09,Lyutikov10}.

Even so, it is intriguing to consider the idea of highly magnetized jets, as the role of the kpc-scale environment in influencing the jet power and structure seems to be undermined in our recent MOJAVE VLA study \citep{Kharb10}. This was concluded primarily from the presence of a strong correlation between the parsec-scale apparent jet speeds and the kpc-scale radio lobe emission, and the lack of a correlation between the parsec-to-kpc-scale jet misalignment and redshift. Magnetically dominated jets are suggested to not be as influenced by interaction with the environment, compared to hydrodynamic jets \citep[e.g.,][]{Lyutikov03}. Moreover, there are indications of highly magnetized jets on scales of several parsecs, through observed rotation measure (RM) gradients transverse to the jets \citep[e.g.,][Hovatta et al., 2012, in prep.]{Kharb09,Croke10}. Bimodal magnetic field structures can be interpreted (although not a unique explanation; \citet{Kharb08b}) as arising from helical/toroidal  fields \citep[e.g.,][]{Lyutikov05,Kharb05,Pushkarev05,Kharb08a}. Helical magnetic fields have also been inferred through the presence of correlated asymmetries in total intensity and fractional polarization in transverse slices across parsec-scale jets \citep[e.g.,][]{Clausen11}. 

As noted earlier, the jet morphology in 0106+013 resembles a ``nose-cone'' from a highly magnetized jet. This morphology is similar to that observed in the quasar 3C\,273 \citep{Bahcall95}, which is another candidate for a ``nose-cone'' and a highly magnetized jet \citep{Clarke86}. 3C\,273 also happens to exhibit the best case of an RM gradient across its parsec-scale jet; RM gradients being a tell-tale signature of large scale helical magnetic fields \citep{Asada02}. The magnetized nature of the jet in 3C\,345 has also been inferred, from jet acceleration on parsec-scales \citep{Vlahakis04}. In short, the shocked jet regions upstream of the radio hot spots, the kpc-scale jet wiggles and a ``nose-cone'' like radio jet in 0106+013, the V-shaped kpc-scale radio structure and jet acceleration on parsec-scales in 3C\,345, are all broadly consistent with instabilities associated with Poynting flux dominated jets. A greater theoretical understanding and sensitive numerical simulations of jets spanning parsec- to kpc-scales are needed, however, to make direct quantitative comparsions. At the observational end, high frequency radio and optical polarimetry of radio jets that can clearly resolve the jets in the transverse direction, and a search for robust Faraday rotation gradients and circular polarization, needs to be carried out. The persistence of the handedness of circular polarization over time could serve as a possible indicator of helical magnetic fields, in addition to being used as a diagnostic of plasma conditions and properties \citep[e.g.,][]{Homan09}. 
}

\acknowledgments
We thank the referee for making useful suggestions which have improved this paper. PK thanks R. Montez, E. Clausen-Brown, M. Lyutikov and M. Huarte Espinosa for enlightening discussions, W. Cui for sharing the SED modeling code, M. Aller for the UMRAO flux densities, and E. Perlman for suggestions on dealing with the PSF spike. BH thanks B. Zitzer for his error analysis contributions. 
Support for this work was provided by the National Aeronautics and Space Administration (NASA) through Chandra Award Number G09-0128X issued by the Chandra X-ray Observatory Center (CXC), which is operated by the Smithsonian Astrophysical Observatory (SAO)  for and on behalf of NASA under contract NAS8-03060. 
Support for Program number HST-GO-11831.04-A was provided by NASA through a grant from the Space Telescope Science Institute, which is operated by the Association of Universities for Research in Astronomy, Incorporated, under NASA contract NAS5-26555.
Support for this work was provided in part by NASA through the SAO contract SV3-73016 to MIT for support of the CXC, which is operated by SAO for and on behalf of NASA under contract NAS8-03060. The National Radio Astronomy Observatory is a facility of the National Science Foundation operated under cooperative agreement by Associated Universities, Inc. 

{\it Facilities:} \facility{VLA}, \facility{HST (ACS)}, \facility{CXO (ACIS)}.

\bibliographystyle{apj}
\bibliography{ms}

\end{document}